\documentclass[twocolumn]{aastex63}
\usepackage[varg]{txfonts}
\usepackage{graphicx}
\begin{document}

\title{Increasing mass-to-flux ratio from the dense core to the protostellar envelope around the Class 0 protostar HH~211}

\author{Hsi-Wei Yen}
\affiliation{Academia Sinica Institute of Astronomy and Astrophysics, 11F of Astro-Math Bldg, 1, Sec. 4, Roosevelt Rd, Taipei 10617, Taiwan}

\author{Patrick Koch}
\affiliation{Academia Sinica Institute of Astronomy and Astrophysics, 11F of Astro-Math Bldg, 1, Sec. 4, Roosevelt Rd, Taipei 10617, Taiwan}

\author{Chin-Fei Lee}
\affiliation{Academia Sinica Institute of Astronomy and Astrophysics, 11F of Astro-Math Bldg, 1, Sec. 4, Roosevelt Rd, Taipei 10617, Taiwan}

\author{Naomi Hirano}
\affiliation{Academia Sinica Institute of Astronomy and Astrophysics, 11F of Astro-Math Bldg, 1, Sec. 4, Roosevelt Rd, Taipei 10617, Taiwan}

\author{Nagayoshi Ohashi}
\affiliation{Academia Sinica Institute of Astronomy and Astrophysics, 11F of Astro-Math Bldg, 1, Sec. 4, Roosevelt Rd, Taipei 10617, Taiwan}

\author{Jinshi Sai}
\affiliation{Academia Sinica Institute of Astronomy and Astrophysics, 11F of Astro-Math Bldg, 1, Sec. 4, Roosevelt Rd, Taipei 10617, Taiwan}

\author{Shigehisa Takakuwa}
\affiliation{Department of Physics and Astronomy, Graduate School of Science and Engineering, Kagoshima University, 1-21-35 Korimoto, Kagoshima, Kagoshima 890-0065, Japan}
\affiliation{Academia Sinica Institute of Astronomy and Astrophysics, 11F of Astro-Math Bldg, 1, Sec. 4, Roosevelt Rd, Taipei 10617, Taiwan}

\author{Ya-Wen Tang}
\affiliation{Academia Sinica Institute of Astronomy and Astrophysics, 11F of Astro-Math Bldg, 1, Sec. 4, Roosevelt Rd, Taipei 10617, Taiwan}

\author{Ken'ichi Tatematsu}
\affiliation{Nobeyama Radio Observatory, National Astronomical Observatory of Japan, National Institutes of Natural Sciences, Nobeyama, Minamimaki, Minamisaku, Nagano 384-1305, Japan}
\affiliation{Department of Astronomical Science, The Graduate University for Advanced Studies, SOKENDAI, 2-21-1 Osawa, Mitaka, Tokyo 181-8588, Japan}

\author{Bo Zhao}
\affiliation{Department of Physics \& Astronomy, McMaster University, Hamilton, ON L8S 4K1, Canada}

\correspondingauthor{Hsi-Wei Yen}
\email{hwyen@asiaa.sinica.edu.tw}

\begin{abstract}
To study transportation of magnetic flux from large to small scales in protostellar sources, we analyzed the Nobeyama 45-m N$_2$H$^+$ (1--0), JCMT 850~$\mu$m polarization, and ALMA C$^{18}$O (2--1) and 1.3~mm and 0.8~mm (polarized) continuum data of the Class 0 protostar HH~211. The magnetic field strength in the dense core on a 0.1 pc scale was estimated with the single-dish line and polarization data using the Davis--Chandrasekhar--Fermi method, and that in the protostellar envelope on a 600 au scale was estimated from the force balance between the gravity and magnetic field tension by analyzing the gas kinematics and magnetic field structures with the ALMA data. Our analysis suggests that from 0.1 pc to 600 au scales, the magnetic field strength increases from 40--107 uG to 0.3--1.2 mG with a scaling relation between the magnetic field strength and density of $B \propto \rho^{0.36\pm0.08}$, and the mass-to-flux ratio increases from 1.2--3.7 to 9.1--32.3. 
The increase in the mass-to-flux ratio could suggest that the magnetic field is partially decoupled from the neutral matter between 0.1 pc and 600 au scales, and hint at efficient ambipolar diffusion in the infalling protostellar envelope in HH~211, which is the dominant non-ideal magnetohydrodynamic effect considering the density on these scales. Thus, our results could support the scenario of efficient ambipolar diffusion enabling the formation of the 20 au Keplerian disk in~HH 211. 
\end{abstract}


\section{Introduction}
Stars form through gravitational collapses of magnetized dense cores \citep{Shu87,Crutcher12}. 
Transportation of the magnetic flux from large to small scales in protostellar sources is one of the fundamental questions in star formation \citep{Paleologou83,Mouschovias85,Nakano86a,Nakano86b}.
In the ideal magnetohydrodynamic (MHD) limit, 
the matter and the magnetic field are coupled in a collapsing dense core.  
The magnetic fields are dragged inward by the collapsing material efficiently,   
and the magnetic flux accumulates in the inner region, resulting in a strong magnetic tension force \citep{Mellon08, Zhao11}. 
Subsequently, the strong magnetic tension force in the protostellar envelope can efficiently slow down gas motions and transfer a significant amount of angular momentum outward, which suppresses formation and growth of a protostellar disk \citep{Li14,Tsukamoto16}.

Non-ideal MHD effects, namely ambipolar diffusion, Hall effect, and Ohmic dissipation, are theoretically expected to enable the magnetic field to partially decouple from neutral matter and play an important role in weakening the magnetic field and redistributing the magnetic flux in a protostellar envelope, hence solving the magnetic flux problem \citep{Li98} and reducing efficiency of magnetic braking \citep{Wurster18,Zhao20}.
The diffusion rates of the non-ideal MHD effects depend on the abundance of charged particles and their momentum transfer with H$_2$ gas. 
The ionization fraction in a protostellar source is determined by its density and cosmic-ray ionization rate. 
Dust also plays an important role in ionization chemistry because ions and electrons can recombine or be absorbed on dust surface.  
Besides, dust grains are main reservoirs of negative charges and thus also contribute to the conductivity.
Therefore, the density, cosmic-ray ionization rate, and dust grain size distribution in a protostellar source can significantly affect the diffusion rates \citep{Dapp12, Padovani14, Marchand16, Zhao16, Dzyurkevich17, Guillet20, Tsukamoto20, Kawasaki22, Tsukamoto22}.
Observationally, it remains unclear at which scale the non-ideal MHD effects become efficient. 
Observational studies of magnetic field strengths and mass-to-flux ratios as a function of spatial scales in protostellar sources in comparison with theoretical simulations should provide clues to the efficiency of the non-ideal MHD effects and magnetic field--matter decoupling in the star formation process \citep{Masson16, Zhao16, Zhao18}. 

On the scales of molecular clouds and dense cores, 
the Davis--Chandrasekhar--Fermi (DCF) method \citep{Davis51,Chandrasekhar53}, which assumes that the perturbation in magnetic field structures is caused by turbulence, is often adopted to estimate magnetic field strengths from the angular dispersion of the magnetic field structures and the turbulent line width \citep{Pattle17, Hwang21, Liu22}.
The uncertainty and accuracy of this method have been studied with numerical simulations \citep{Ostriker01, Padoan01, Liu21, Chen22}. 
The DCF method has also been applied to the polarization data of protostellar envelopes on a scale smaller than 1000 au \citep{Girart06, Hull17, Maury18, Kwon19}. 
Nevertheless, since protostellar envelopes on a 1000 au scale are typically dynamically infalling, it is still uncertain if the assumption of the DCF method is valid on the envelope scale and observed angular dispersions of the magnetic field structures are caused by turbulence. 
An alternative method to estimate the magnetic field strength in protostellar envelopes on a 1000 au scale is to compare observed infalling velocity with expected free-fall velocity \citep{Aso15,Sai22}. 
On the assumption that the observed ratio of infalling to expected free-fall velocities is related to the balance between the gravitational and magnetic field tension forces in a protostellar envelope, the magnetic field strength can be estimated from the infalling velocity, enclosed mass, and magnetic field structures in an protostellar envelope. 

HH~211 is a Class 0 protostar in the Perseus star-forming region at a distance of 320 pc \citep{Ortiz-Leon18}.
HH~211 is an edge-on system with an inclination angle of 81$\arcdeg$ \citep{Jhan16}, 
and it launches collimated jets and outflows along the axis with a position angle of 116$\arcdeg$ \citep{Gueth99, Lee09, Jhan16, Jhan21}. 
HH~211 is surrounded by an infalling and rotating protostellar envelope \citep{Tanner11, Lee19} and a Keplerian disk with a radius of 20 au \citep{Lee18}.
Its protostellar mass has been estimated to be 0.08 $M_\sun$ from the Keplerian rotation observed in the SO emission at a resolution of 0\farcs06 with the 
Atacama Large Millimeter/submillimeter Array \citep[ALMA;][]{Lee18}.
The magnetic field structures on a 0.1 pc scale around HH~211 have been revealed with polarimetric observations with the James Clerk Maxwell Telescope \citep[JCMT;][]{Matthews09, Yen21}. 
The ALMA polarimetric observations have detected pinched magnetic fields in the protostellar envelope on a 600 au scale around HH~211, possibly due to the infalling motion in the envelope \citep{Lee19}. 

In this work, we make use of the archival data of the magnetic field structures in the dense core and protostellar envelope of HH~211 together with the molecular-line data to estimate the magnetic field strengths on scales of 0.1 pc and 600 au in HH~211.
In the present paper, we introduce the data used in this study in Section \ref{data}, 
and describe our analysis to estimate magnetic field strengths with the DCF method and from the force balance between the gravity and the magnetic field tension in Section \ref{analysis}. 
From the change in the magnetic field strengths from the dense core to protostellar envelope scales, 
we discuss the possible decoupling between the magnetic field and the matter in HH~211 and the implication on the non-ideal MHD effects in the star formation process, in Section \ref{discussion}. 

\section{Observations}\label{data}
\subsection{Nobeyama 45-m N$_2$H$^+$ (1--0) data}
Observations of HH~211 in the N$_2$H$^+$ (1--0) line were conducted as a part of on-the-fly (OTF) mapping of $7\farcm5 \times 5\farcm4$ area in the IC~348 region with the 45-m radio telescope of the Nobeyama Radio Observatory between 2021 December 6 and 24 (program ID: CG211006). 
The half-power beam width of the 45-m telescope is 17\farcs8 at 93 GHz. 
The SMA45 and FOREST were adopted as the receiver backend and frontend \citep{Kamazaki12, Minamidani16}, respectively. 
The channel width and band width of the spectral window for the N$_2$H$^+$ (1--0) line were 7.63 kHz and 15.63 MHz, respectively.
OTF mapping was performed along right ascension and declination, 
and pointing was checked with nearby SiO masers every 1--1.5 hours.

The data were reduced with the NOSTAR program \citep{Sawada08}. 
Linear baselines were first subtracted from the spectra. 
Then the data were convolved with a Bessel--Gaussian function and gridded to generate maps with a pixel size of 6$\arcsec$ and a channel width of 0.1 km s$^{-1}$, 
and the spectra taken with scans along right ascension and declination were reduced separately in this step. 
This results in an effective angular resolution of the maps of 19\farcs5. 
Finally the maps from the scans along the two orthogonal directions were basket-weaved \citep{Emerson88}.
The main beam efficiency\footnote{\url{https://www.nro.nao.ac.jp/~nro45mrt/html/prop/eff/eff_latest.html}} at 93 GHz was estimated to be 46\%. 
The pixel values in the maps were divided by the main beam efficiency to produce the maps in the unit of the main beam temperature.
The noise level of the maps in main beam temperature is 0.38 K.

\subsection{JCMT polarized 850~$\mu$m continuum data}
We retrieved the 850 $\mu$m continuum data of the IC~348 region with a field of view of 14$\arcmin$ in diameter taken with the POL-2 instrument on JCMT, which has an angular resolution of 14\farcs6, from the archive (Proposal ID: M17BL011 and M17BP058). 
The observations were carried out between 2017 July 7 and September 23 and between 2019 October 5 and 2020 February 25. 
The data were reduced using the software {\it Starlink} \citep{Currie14} and the task {\it pol2map} of the version of 2021A following the standard procedure\footnote{\url{https://www.eaobservatory.org/jcmt/science/reductionanalysis-tutorials/pol-2-dr-tutorial-1/}}. 
The data were first reduced with the default pixel size of 4$\arcsec$, 
and the Stokes {\it IQU} maps were binned to a pixel size of 12$\arcsec$, comparable to the angular resolution, to extract polarization detections. 
The polarization intensity was debiased. 
Our detection criteria of the polarized emission are signal-to-noise ratios (S/N) of Stokes {\it I} intensities higher than five, S/N of polarized intensities higher than two, and polarization percentages lower than 30\%. 
In this paper, we present the 850 $\mu$m continuum map and polarization data in the sub region of 2\farcm5 by 2\farcm5 centered at HH~211 in the IC~348 region observed with JCMT.
There are 64 detections within a radius of 1$\arcmin$ around HH~211, and all of them have high S/Ns above 3$\sigma$.

The 850 $\mu$m flux conversion factor (FCF) of SCUBA-2 is 516$\pm$42 Jy beam$^{-1}$ pW$^{-1}$ for the data taken in 2017 and is 495$\pm$32 Jy beam$^{-1}$ pW$^{-1}$ for the data taken in 2019--2020 \citep{Mairs21}, 
and the POL-2 instrument has a FCF a factor of 1.35 higher than SCUBA-2. 
Thus, we adopted a single FCF of 679 Jy beam$^{-1}$ pW$^{-1}$ to convert the Stokes {\it I} intensity of our co-added POL-2 map.
In the following analysis, the Stokes {\it I} map with a pixel size of 4$\arcsec$ was used to measure the 850 $\mu$m continuum flux. 

\subsection{ALMA 1.3 mm and 0.8 mm data}
We retrieved the raw visibility data of the 0.8 mm and 1.3 mm continuum and C$^{18}$O (2--1) emission in HH~211 observed with ALMA from the archive (Project code: 2016.1.00017.S and 2017.1.01078.S).
The details of the observations have been introduced in \citet{Lee19} and \citet{Tychoniec20}. 
The data were reduced using the calibration scripts from the archive with the corresponding CASA versions. 
Then we performed self calibration on the phase of the continuum data, 
and the solutions were applied to the molecular-line data.
The combination of these data sets has baseline lengths from 15 m to 3.7 km for the 1.3 mm continuum and C$^{18}$O (2--1) and from 19 m to 3.1 km for the 0.8 mm continuum. 
Our analysis of the ALMA data was performed in the visibility domain. 
Images were generated for visualization of the results of our analysis. 

The 0.8 mm and 1.3 mm continuum and C$^{18}$O (2--1)  images were generated using the task {\it tclean} of the CASA version of 6.4 with the Briggs weighting of a robust parameter of 0.5, resulting in the synthesized beams of 0\farcs18$\times$0\farcs11 ($-5\arcdeg$), 0\farcs16$\times$0\farcs1 ($-22\arcdeg$), and 0\farcs2$\times$0\farcs13 ($-21\arcdeg$) and the noise levels of 0.1, 0.02, and 1.6 mJy beam$^{-1}$, respectively.
The C$^{18}$O (2--1) image cube has a channel width of 0.17 km s$^{-1}$.
In addition, we obtained the table of 1.3 mm polarization detections from \citet{Lee19}, including positions, polarized intensities, and polarization angles.
In \citet{Lee19}, the polarized intensity is debiased, and the criterion of a polarization detection is the polarized intensity above 2.5$\sigma$.

\section{Analysis and results}\label{analysis}
\subsection{Dense core on a 0.1 pc scale}
\subsubsection{Core identification and velocity}\label{sec_45m_n2hp}

\begin{figure*}
\centering
\includegraphics[width=\textwidth]{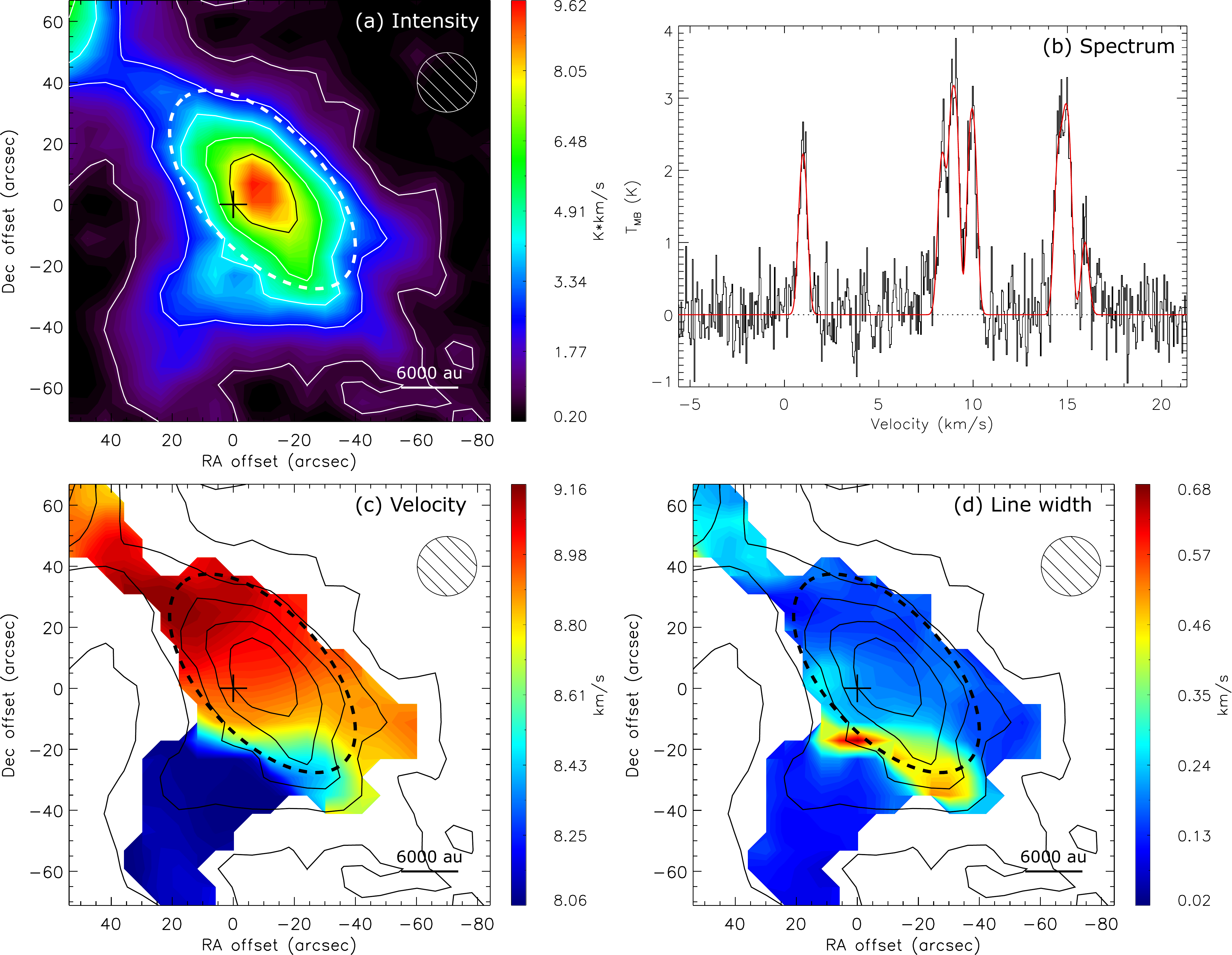}
\caption{(a) Total integrated intensity map of the N$_2$H$^+$ (1--0) emission in HH~211 observed with the Nobeyama 45-m telescope. Contours are from 10\% in steps of 20\% of the peak intensity, which is 9.62 K km s$^{-1}$. The noise level is 0.2 K km s$^{-1}$. (b) N$_2$H$^+$ (1--0) spectrum at the protostellar position (black). A red curve presents the hyperfine fitting result. The noise level is 0.38 K. (c) \& (d) Centroid velocity and line width measured from the fitting to the hyperfine structures of the N$_2$H$^+$ (1--0) emission, respectively, where the line width is defined as 1$\sigma$ width of the fitted Gaussian profile. Crosses and hatched circles denote the protostellar position and the angular resolution, respectively. Dashed open ellipses delineate the dense core associated with HH~211 identified in the N$_2$H$^+$ (1--0) emission.}\label{45m_n2hp}
\end{figure*}

Figure~\ref{45m_n2hp}a presents the total integrated intensity map of all hyperfine components of the N$_2$H$^+$ (1--0) emission obtained with the Nobeyama 45-m observations.
The N$_2$H$^+$ emission around HH~211 is elongated along the northeast--southwest direction, and the emission in the south is further extended toward the southeast. 
The extension toward the northeast is connected to other dense cores in the IC~348 region \citep{Walker-Smith14}.
We have performed fitting to the seven hyperfine components of N$_2$H$^+$ (1--0) on the assumption of the local thermal equilibrium \citep[LTE;][]{Mangum15}.  
There are four free parameters in our fitting to the N$_2$H$^+$ (1--0) line, the excitation temperature, optical depth, centroid velocity, and line width. 
Our fitting result of the N$_2$H$^+$ (1--0) spectrum at the protostellar position is presented in Fig.~\ref{45m_n2hp}b as an example.
The protostellar position is adopted to be the peak position of the 0.8 mm continuum emission, 3$^{\rm h}$43$^{\rm m}$56\fs81, 32$\arcdeg$00$\arcmin$50$\farcs$2, measured from the ALMA data \citep[Section \ref{sec_alma_cont};][]{Lee18}.
Figure~\ref{45m_n2hp}c and d present the maps of the centroid velocity and  line width measured with the hyperfine fitting to the N$_2$H$^+$ (1--0) line, 
where the line width is defined as 1$\sigma$ width of the fitted Gaussian profile.  
A velocity gradient from the southwest (blueshifted) to northeast (redshifted) around HH~211 is seen. 
The direction of this velocity gradient is consistent with that observed on a smaller scale of 20$\arcsec$ in the NH$_3$ emission with VLA \citep{Tanner11}. 
In our N$_2$H$^+$ velocity map, there is an additional velocity gradient toward the southeast, where the emission is even more blueshifted.   
The line width increases to be more than 0.5 km s$^{-1}$ in the intersection of the two velocity gradients, 
while in the other regions the line widths are mostly 0.1--0.2 km s$^{-1}$. 
These results suggest that there are likely two velocity components in this region, 
and the measured line width in the intersection increases because the two velocity components are blended. 

We applied the {\it dendrogram} algorithm\footnote{https://dendrograms.readthedocs.io/en/stable/index.html} to the N$_2$H$^+$ velocity channel maps to identify the structures associated with HH~211. 
Because there are multiple hyperfine components in the N$_2$H$^+$ velocity channel maps, 
the {\it dendrogram} algorithm cannot be directly applied.
To avoid confusion due to the hyperfine structures of N$_2$H$^+$ (1--0), 
we first subtracted the best-fit N$_2$H$^+$ (1--0) spectra from the data and generated residual velocity channel maps. 
We confirmed that the maximum residual is 3.1$\sigma$ and the standard deviation of the residuals is the same as the noise level.
Thus, all the N$_2$H$^+$ emission was subtracted. 
Then we added the best fits of the strongest hyperfine component at 93.173764 GHz to the residual maps, 
so only one hyperfine component left in the N$_2$H$^+$ velocity channel maps. 
We applied {\it dendrogram} to these processed velocity channel maps. 
Open dashed ellipses in Fig.~\ref{45m_n2hp} delineate the dense core associated with HH~211 identified with {\it dendrogram}, which has a size\footnote{The semi-major and -minor axes are defined as twice the intensity-weighted second moments of the identified structure \citep{Rosolowsky08}.} of $80\arcsec \times 40\arcsec$ with a position angle of the major axis of 48$\arcdeg$, and is centered at 3$^{\rm h}$43$^{\rm m}$56\fs1, 32$\arcdeg$00$\arcmin$55$\arcsec$ and $V_{\rm LSR}$ of 9.0 km s$^{-1}$. 
This area is adopted to measure the properties of the dense core associated with HH~211 in our analysis. 
The velocity gradient in the dense core on a 60$\arcsec$ scale is measured to be 4.6 km s$^{-1}$ pc$^{-1}$, 
and this magnitude is approximately 80\% of that of the NH$_3$ emission in the inner envelope on a 20$\arcsec$ scale observed at a 4$\arcsec$ resolution  \citep{Tanner11}.

From the line width map (Fig.~\ref{45m_n2hp}d), the mean 1$\sigma$ line width ($\delta v_{\rm ob}$) of the dense core associated with HH~211 is measured  to be 0.18 km s$^{-1}$.
The uncertainty of the line width measurement in the dense core due to the noise is 0.01--0.02 km s$^{-1}$.
We note that the area of the  identified dense core slightly covers the region with larger line widths.
Nevertheless, the measured mean line width is not sensitive to the exact area of the dense core. 
Even if the core radius is twice smaller, the mean line width would only change by 10\%, which is comparable to the uncertainty in the line width due to the noise.
We assume the mean kinetic temperature of the dense core to be 16$\pm$0.2 K,  identical to the dust temperature measured by fitting the spectral energy distributions at a resolution of 36$\arcsec$ from the Herschel data by \citet{Zari16}, 
and the thermal line width of N$_2$H$^+$ ($\delta v_{\rm th}$) in the dense core is estimated to be 0.09 km s$^{-1}$.
Thus, the mean non-thermal line width ($\delta v_{\rm nt}$) of the dense core associated with HH~211 is estimated to be 0.15 km s$^{-1}$, as $\delta v_{\rm nt} = \sqrt{{\delta v_{\rm ob}}^2-{\delta v_{\rm th}}^2}$, 
which is in the range of $\delta v_{\rm nt}$ of 0.03 to 0.46 km s$^{-1}$ within a 20$\arcsec$ scale estimated with the VLA NH$_3$ observations \citep{Tanner11}.

\subsubsection{Magnetic field structure and strength}

\begin{figure*}
\centering
\includegraphics[width=\textwidth]{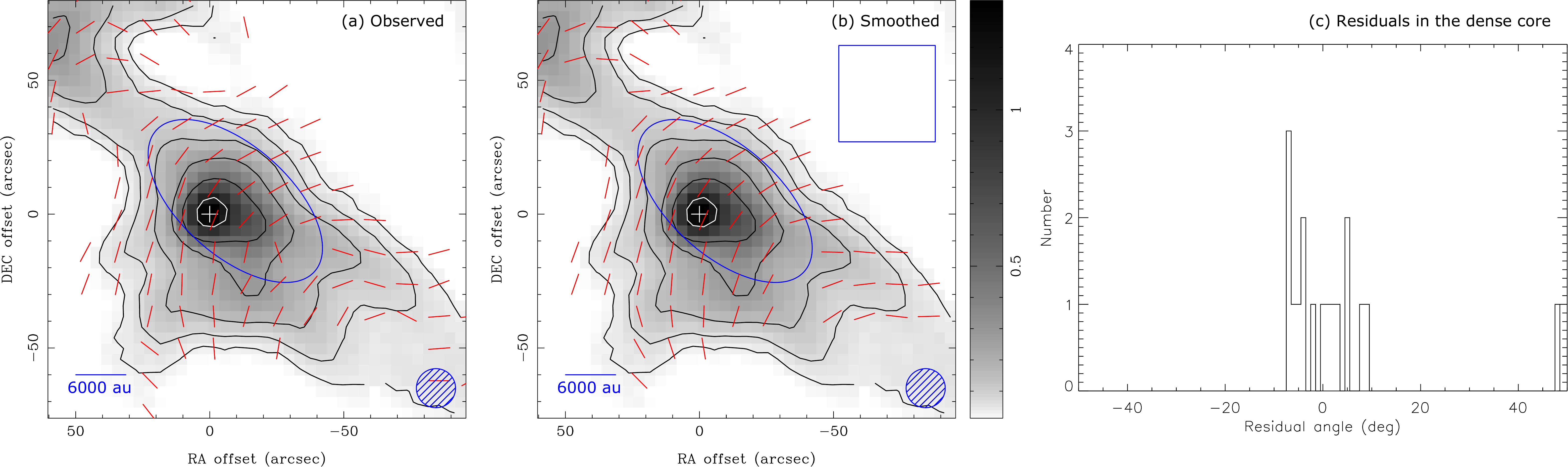}
\caption{850 $\mu$m continuum map (gray scale and contours; in units of Jy beam$^{-1}$) on a 0.1~pc scale around HH~211 observed with JCMT. The segments show (a) the magnetic field orientations inferred by rotating the observed polarization orientations by 90$\arcdeg$ and (b) the smoothed magnetic field structures after averaging over every 3 by 3 detections (36$\arcsec \times 36\arcsec$), corresponding to the scale delineated by an open blue box.  
The separation of two adjacent polarization detections is 12$\arcsec$.
Hatched circles present the angular resolution of the JCMT observations of 14\farcs6. 
Contour levels are from 2.5\% of the peak intensity and in steps of a factor of two, where the peak intensity is 1.35 Jy beam$^{-1}$. 
The noise level is 3 mJy beam$^{-1}$. 
Crosses denote the protostellar position. Open ellipses delineate the dense core associated with HH~211 identified in the N$_2$H$^+$ (1--0) emission. 
Panel (c) present residual angles in the dense core after subtracting the smoothed magnetic field structures from the observed magnetic field orientations. There are 17 detections of the magnetic-field orientations within the area of the dense core identified in the N$_2$H$^+$ (1--0) emission.}\label{jcmt_pol2}
\end{figure*}

\begin{figure}
\center
\includegraphics[width=0.45\textwidth]{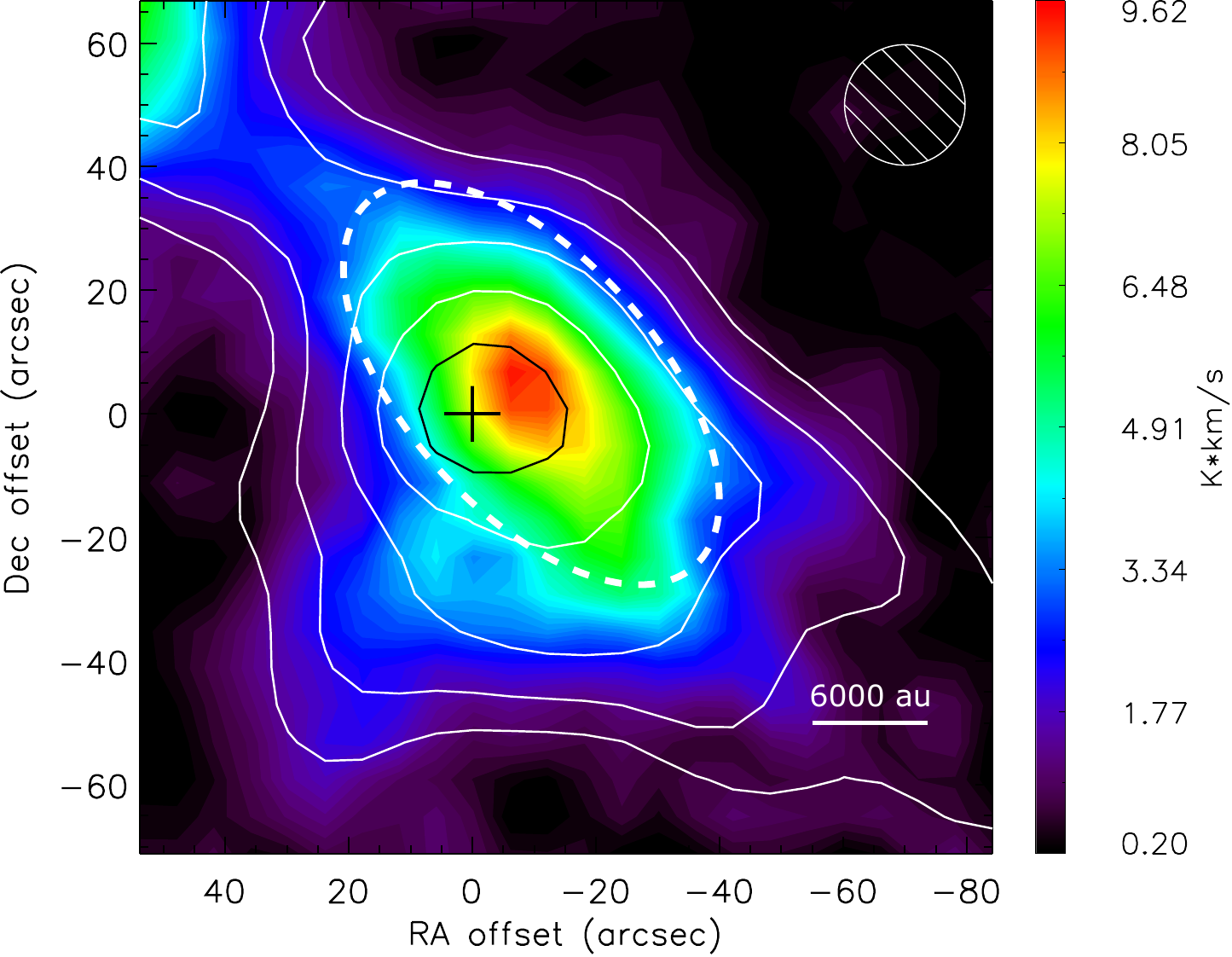}
\caption{Comparison of the intensity distributions of the N$_2$H$^+$ (1--0; color) and 850 $\mu$m continuum (contours) emission. The N$_2$H$^+$ map is the same as that in Fig.~\ref{45m_n2hp}a. The 850 $\mu$m continuum map was convolved to have the same resolution as that of the N$_2$H$^+$ map. Contour levels are from 2\% of the peak intensity and in steps of a factor of two, where the peak intensity is 1.73 Jy beam$^{-1}$. 
 A cross and a hatched circle denote the protostellar position and the angular resolution, respectively. A dashed open ellipse delineates the dense core associated with HH~211 identified in the N$_2$H$^+$ (1--0) emission.}\label{n2hp_850um}
\end{figure}

Figure~\ref{jcmt_pol2}a presents the 850 $\mu$m continuum map and the magnetic field structures on a 0.1~pc scale around HH~211 observed with JCMT.  
The intensity distribution of the 850 $\mu$m continuum emission is elongated along the northeast--southwest direction with an extension toward the southeast. 
Figure~\ref{n2hp_850um} compares the intensity distributions of the N$_2$H$^+$ (1--0) and 850 $\mu$m continuum emission. 
The 850 $\mu$m continuum map in Fig.~\ref{n2hp_850um} was convolved to have a resultant angular resolution the same as that of the N$_2$H$^+$ map for the comparison. 
The similarity between the intensity distributions of the N$_2$H$^+$ (1--0) and 850 $\mu$m continuum emission suggests that the N$_2$H$^+$ (1--0) and 850 $\mu$m continuum emission likely traces similar volume, although there is an offset  of 7$\arcsec$ between the 850 $\mu$m continuum and N$_2$H$^+$ (1--0) intensity peaks, which is less than two pixels in the N$_2$H$^+$ map. 
We adopted the dust temperature of 16 K in the dense core measured from the Herschel data to estimate the core mass and column density \citep{Zari16}.  
We first computed the optical depth of the 850 $\mu$m continuum emission in the dense core using the Stokes {\it I} map obtained with POL-2, which ranges from 5$\times$10$^{-4}$ to 6$\times$10$^{-3}$ with a mean value of 2$\times$10$^{-3}$. 
The dust absorption coefficient at 850 $\mu$m was assumed to be 0.022 cm$^2$ g$^{-1}$ from the opacity table of dust with thick ice mantles in \citet{Ossenkopf94}, including a gas-to-dust mass ratio of 100.
The mass, mean column density ($\Sigma$), and mean density\footnote{We assumed that the dense core is an ellipsoid with principal axes of 80$\arcsec$, 40$\arcsec$, and 40$\arcsec$--80$\arcsec$, where the principal axis along the line of sight is assumed to be between the major and minor axes on the plane of the sky (40$\arcsec$--80$\arcsec$), and the mean density was estimated by dividing the core mass by the volume of the ellipsoid.} ($\rho$) of the dense core were estimated to be 2.5 $M_\sun$, 0.09 g cm$^{-2}$, and (3.3--6.6)$\times$10$^{-19}$ g cm$^{-3}$, respectively, from the optical depths and the dust absorption coefficient.
Their uncertainties due to the noise in the 850 $\mu$m continuum data and the dust temperature measurements are 2\%, and the uncertainty in the FCF is 6\%--8\%. 
Thus, in the following analysis, we assume the uncertainty in the mass and column density of the dense core to be 10\%, and there is an uncertainty of a factor of two in the mean density because of the unknown length along the line of sight of the dense core.

The direction of the magnetic fields around HH~211 is from the south to the northwest (Fig~\ref{jcmt_pol2}a), 
and the overall morphology is consistent with that observed with SCUPOL \citep{Matthews09}.
The mean orientation of the magnetic fields in the dense core associated with HH~211 has a position angle of 145$\arcdeg$, 
which is derived from the mean Stokes {\it Q} and {\it U} intensities in the dense core, 
and the mean uncertainty of the individual observed magnetic field orientations in the dense core is 2$\arcdeg$.
The large-scale magnetic field structures are first removed to measure the angular dispersion of the magnetic field orientations in the dense core associated with HH~211.
We adopt the method similar to unsharp masking to extract the structures of the large-scale magnetic fields, as applied to other POL-2 data \citep{Pattle17,Wang19}. 
The smoothing scale for unsharp masking should be larger than the coherence scale of the perturbed magnetic fields by the turbulence and smaller than the scale of the variation of the large-scale magnetic fields.
If a smoothing scale is smaller than the coherence scale or larger than the variation scale, the angular dispersion would be under- or over-estimated after removing the extracted large-scale magnetic field structures \citep{Chen22}.
Our purpose is to estimate the angular dispersion of the magnetic fields in the dense core with a size of $80\arcsec \times 40\arcsec$, 
and it is also seen that the orientation of the large-scale magnetic fields varies over the scale of the dense core (Fig~\ref{jcmt_pol2}a).
However, it is not straightforward to quantify the coherence and variation scales.
Thus, we have tried two different smoothing scales, 3 by 3 pixels ($36\arcsec \times 36\arcsec$), where the size of the 3 pixels approximately corresponds to the size of the dense core along the minor axis, and 5 by 5 pixels ($60\arcsec \times 60\arcsec$), the geometric mean diameter of the dense core.
The extracted large-scale magnetic field structures with a smoothing scale of 3 by 3 pixels and a histogram of the residual angles after subtracting these large-scale magnetic field structures are shown in Fig.~\ref{jcmt_pol2}b and c, respectively, as an example. 
The standard deviation of the residual angles is computed to be 12$\arcdeg$--15$\arcdeg$ with the smoothing scales of 3 by 3 and 5 by 5 pixels, 
and they are adopted as the angular dispersion of the magnetic field orientations in the dense core.
We have also estimated the angular dispersion with the method using the structure function of the magnetic field orientations, as introduced in \citet{Hildebrand09}, 
and obtained a consistent result of the angular dispersion of 12$\arcdeg$.

With the angular dispersion ($\delta\theta_{\rm B}$) of 12$\arcdeg$--15$\arcdeg$ and the density ($\rho$) of (3.3--6.6)$\times$10$^{-19}$ g cm$^{-3}$, 
the magnetic field strength projected on the plane of the sky ($B_{\rm pos}$) in the dense core associated with HH~211 is estimated to be 56--107 $\mu$G with the DCF method \citep{Davis51,Chandrasekhar53} as, 
\begin{equation}\label{dcf}
B_{\rm pos} = \xi \sqrt{4 \pi \rho} \frac{\delta V_{\rm nt}}{\delta\theta_{\rm B}},
\end{equation}
where $\xi$ is adopted to be 0.5 to correct for inhomogeneous and complex magnetic field and density structures along the line of sight \citep{Ostriker01,Chen22}.
On the other hand, \citet{Skalidis21} propose an alternative equation to estimate $B_{\rm pos}$ from $\delta\theta_{\rm B}$ and $\rho$ assuming molecular clouds with compressible and magnetized turbulence, 
\begin{equation}
B_{\rm pos} = \sqrt{4 \pi \rho} \frac{\delta V_{\rm nt}}{\sqrt{2\delta\theta_{\rm B}}}.
\end{equation}
With this equation, the magnetic field strength in the dense core is estimated to be 40--69 $\mu$G. 
We note that these two methods are based on different assumptions on the turbulence properties, and may be applicable in different circumstances. 
As shown in a recent study by \citet{Li22}, both methods provide reasonable estimates of magnetic field strength in their numerical simulations of filaments formed in supersonic turbulent and magnetized molecular clouds, regardless of the different assumptions. 
We therefore applied both methods for comparison.
With the overall range of the estimated magnetic field strengths (40--107 $\mu$G), 
the dimensionless mass-to-flux ratio ($\lambda$) of the dense core associated with HH~211 is estimated to be 1.2 to 3.7 as, 
\begin{equation}\label{mtf}
\lambda = 2\pi\sqrt{G}\frac{\Sigma}{B_{\rm pos}}, 
\end{equation}
where $G$ is the gravitational constant and $\Sigma$ is the column density. 
The ranges of the estimated magnetic field strengths and mass-to-flux ratios here include the uncertainties in the angular dispersion of the magnetic field orientations and the density of the dense core, but does not include correction for the unknown inclination of the magnetic field. 
In addition, the values are subject to the assumed dust absorption coefficient at 850 $\mu$m.
Increasing the dust absorption coefficient by 50\% leads to a 50\% decrease in $\rho$ and $\Sigma$ and a 20\%--30\% decrease in $\lambda$. 

\subsection{Protostellar envelope on a 600 au scale}
\subsubsection{Density and temperature profiles}\label{sec_alma_cont}

\begin{figure*}
\centering
\includegraphics[width=\textwidth]{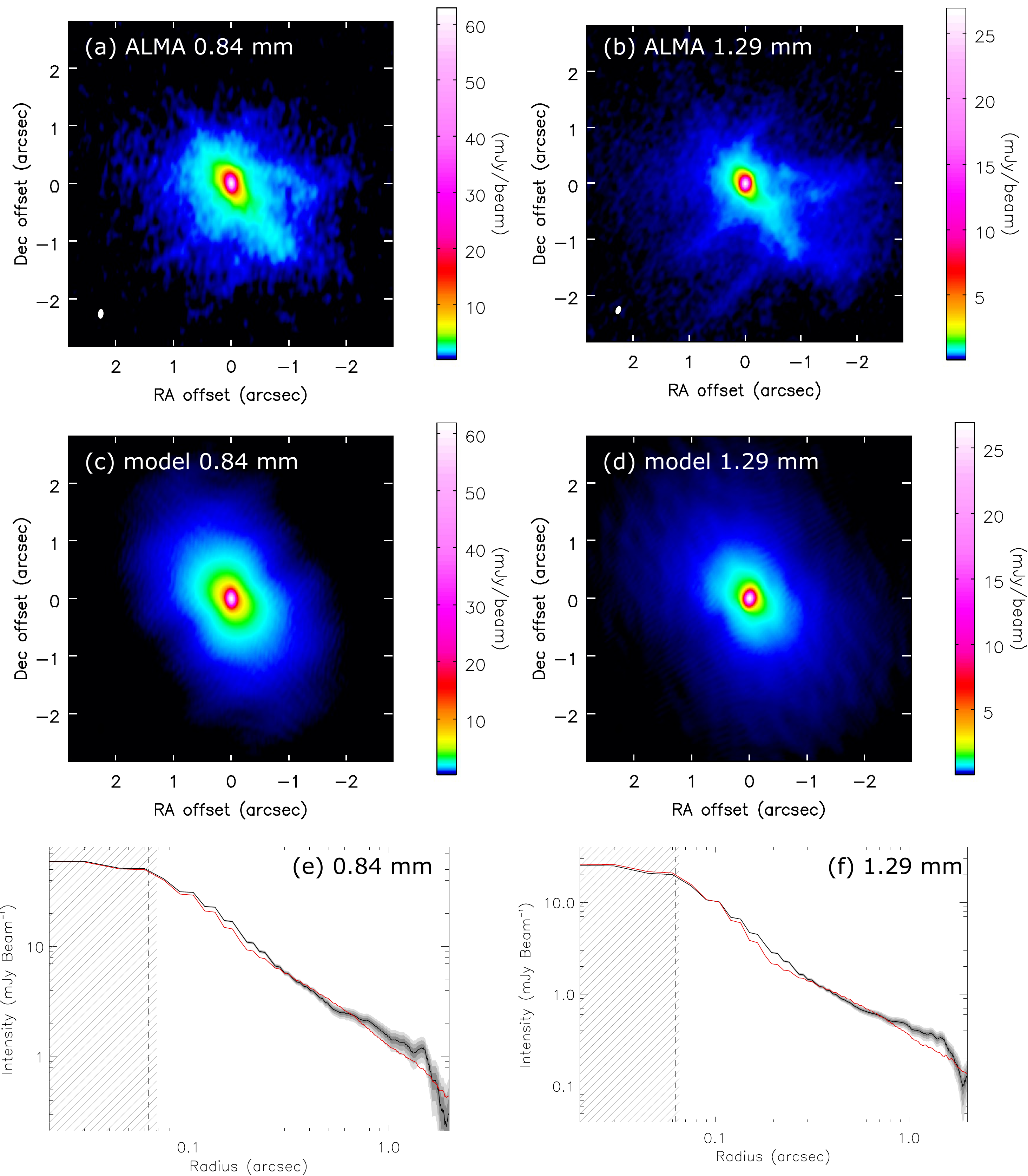}
\caption{(a) \& (b) ALMA 0.8 mm and 1.3 mm continuum images of HH~211. White filled ellipses present the sizes of the synthesized beams of 0\farcs18$\times$0\farcs11 (58 au $\times$ 35 au) at 0.8 mm and 0\farcs16$\times$0\farcs1 (51 au $\times$ 32 au) at 1.3mm. (c) \& (d) Synthetic 0.8 mm and 1.3 mm continuum images from our best-fit model. (e) \& (f) Radial profile of the 0.8 mm and 1.3 mm continuum intensity along the midplane of the protostellar envelope around HH~211 (black curves), where the northeastern and southwestern sides are averaged. Dark to light gray shaded regions show the 1$\sigma$ to 3$\sigma$ uncertainties. Red curves present the intensity profiles extracted from our model images. Hatched regions and vertical dashed lines denote half of the synthesized beams and the disk radius, respectively.}\label{alma_cont}
\end{figure*}

Figure~\ref{alma_cont} presents the 0.8 mm and 1.3 mm continuum images of HH~211 obtained with the ALMA observations. 
The 0.8 mm and 1.3 mm continuum emission is elongated along the northeast--southwest direction with a peak position of 3$^{\rm h}$43$^{\rm m}$56\fs81, 32$\arcdeg$00$\arcmin$50$\farcs$2. 
This peak position is adopted as the protostellar position in the present study. 
The direction of the elongation of the continuum emission on a scale of 600 au ($\sim$2\arcsec) is similar to that of the large-scale structure observed with the single-dish telescopes (Fig.~\ref{n2hp_850um}). 
These ALMA 0.8 mm and 1.3 mm continuum maps have been discussed in detail in \citet{Lee19}.
The continuum emission observed with ALMA most likely traces the flattened protostellar envelope and an embedded disk in HH~211, 
and the position angle of the major axis of the continuum emission has been measured to be 36$\arcdeg$ \citep{Lee19}, 
which is adopted as the midplane of the protostellar envelope in the present study.
The peak brightness temperatures are measured to be 41$\pm$1 K at 1.3 mm and 35$\pm$1 K at 0.8 mm.
The comparable peak brightness temperatures at the two wavelengths suggest that the continuum emission is most likely optically thick at the center.
The disk embedded in the protostellar envelope has been resolved with ALMA at a higher resolution of 0\farcs06, and it has a radius of 20 au (0\farcs06) and a height of 10 au (0\farcs03) \citep{Lee18}. 
The position angle of the major axis of the disk is 28$\arcdeg$, which is perpendicular to the outflow axis of 116$\arcdeg$ and is slightly misaligned with that of the protostellar envelope on a 600 au scale by 8$\arcdeg$ \citep{Lee18,Lee19}.
This disk is not resolved in the data analyzed in the present paper, which have a resolution two to three times larger than the disk radius.
 
We have constructed models of a protostellar envelope with an embedded disk and performed the fitting to the continuum visibility data. 
In our model, the disk radius ($r_{\rm d}$) and height ($h_{\rm d}$) are fixed to be 20 au and 10 au, respectively, 
and the disk is assumed to have uniform density ($\rho_{\rm disk}$) and temperature ($T_{\rm disk}$) because the disk is not resolved with our data.
Our model envelope is assumed to have power-law density ($\rho_{\rm env}$) and temperature ($T_{\rm env}$) profiles in spherical coordinates as, 
\begin{eqnarray}
\rho_{\rm env}(r,\theta) & = & \rho_0 (\frac{r}{r_0})^p \sin^f\theta, \\
T_{\rm env}(r) & = & T_0 (\frac{r}{r_0})^{q}, 
\end{eqnarray}
where $r_0$ is set to be 100 au, $\theta$ is the polar angle, and $\sin^f\theta$ determines the flattenness of the envelope \citep{Brinch07}. 
The power-law index of the temperature profile is fixed to be $-0.4$, a typical value in protostellar sources \citep {Shirley02}, because the density and temperature profiles are highly degenerated and cannot be constrained simultaneously with our data at 0.8 mm and 1.3 mm. 
Thus, there are six free parameters in our model, which are $\rho_0$, $p$, $f$, $T_0$, $\rho_{\rm disk}$, and $T_{\rm disk}$.

Then we calculated radiative transfer using the Simulation Package for Astronomical Radiative Xfer\footnote{\url{https://sparx.tiara.sinica.edu.tw}}(SPARX) and generated model 0.8 mm and 1.3 mm continuum images. 
The opacity table of dust with thick ice mantles in \citet{Ossenkopf94} was adopted in our calculations with the absorption coefficients of 0.023 and 0.01 cm$^2$ g$^{-1}$ at 0.8 mm and 1.3 mm, respectively, the same as for our analysis of mass of the dense core.  
The inclination angle and the position angle of the major axis of the protostellar envelope were adopted to be 81$\arcdeg$ and 36$\arcdeg$ in our model, the same as those in HH~211 \citep{Lee19}, 
and the disk was assumed to have the same orientation as the envelope for simplicity because the disk was not resolved in our data.
The model images were Fourier transformed and sampled with the identical {\it uv} coverages of the ALMA data using the CASA task {\it ft}, 
and the model visibilities were subtracted from the observed visibilities using the CASA task {\it uvsub} to compute the residuals. 
The minimization of the residuals was conducted in the visibility domain using the python package {\it emcee}\footnote{\url{https://emcee.readthedocs.io}} \citep{Foreman-Mackey13}. 

The best-fit parameters are listed in Table~\ref{bestfit}, 
and the uncertainties in these parameters due to the noise in the data are 1\%.
The corner plots of our fitting showing the correlations between the fitting parameters are presented in Appendix~\ref{app_mcmc}.
The synthetic 0.8 mm and 1.3 mm continuum images generated from the visibility data of our best-fit model are shown in Fig.~\ref{alma_cont}c and d.
We note that detailed structures are seen in the observed continuum images, and they cannot be reproduced with our simple axisymmetric model. 
Nevertheless, our model can well explain the observed intensity profiles along the midplane of the protostellar envelope around HH~211 (Fig.~\ref{alma_cont}e and f).

In our best-fit model, the temperature is 20 K at a radius of 100 au and decreases to be 10 K at a radius of 500 au. 
This is consistent with the temperature ranging from 10 K to 20 K on a 1000 au scale estimated from the NH$_3$ data at a resolution of 4$\arcsec$ by \citet{Tanner11}.
The disk mass is 0.09 $M_\sun$ and the envelope mass within a radius of 300 au is 0.06 $M_\sun$ in our best-fit model. 
Our estimated disk mass is ten times larger than the previous estimate\footnote{The difference in the adopted distances was corrected.} of 3--9 $M_{\rm Jup}$ using the 0.8 mm continuum data \citep{Lee18}. 
Our disk temperature is similar to that adopted in \citet{Lee18} within 10\%, and 
the dust absorption coefficient is consistent with theirs within a factor of two. 
The difference in the estimated disk masses is because the continuum emission is assumed to be optically thin in \citet{Lee18}, but our radiative transfer calculations suggest that the continuum emission in the disk is optically thick with an optical depth of 3--7 at 0.8 mm. 
As discussed in \citet{Lee18}, if a larger dust absorption coefficient is adopted in the calculation, the estimated disk mass becomes smaller.
For example, if we adopt the absorption coefficient of 0.035 cm$^{2}$ g$^{-1}$ at 0.8 mm from \citet{Beckwith90}, 
our estimated disk mass would become 0.06 $M_\sun$. 
We also note that our estimated disk mass is comparable to the protostellar mass of 0.08 $M_\sun$ estimated from the Keplerian rotation observed in the SO emission at a high resolution of 0\farcs06 \citep{Lee18}.  
Comparable disk and protostellar masses are also seen at early evolutionary stages ($<$0.1 Myr) when the protostellar mass is low ($<$0.1--0.2 $M_\sun$) in some theoretical simulations \citep{Hennebelle20,Tsukamoto20}.

\begin{deluxetable*}{ccc}
\tablecaption{Best-fit model of a protostellar envelope with an embedded disk to the ALMA data}
\centering
\tablehead{Parameter & Description & Value}
\startdata
& Fitting to the 0.8 mm and 1.3 mm continuum data & \\
\hline
$\rho_0$ & Density at a radius of 100 au in the envelope & $6.2\times10^8$ cm$^{-3}$\\
$p$ & Power-law index of the density profile in the envelope & $-1.5$ \\
$f$ & Flattenness of the envelope & 1.0 \\
$T_0$ & Temperature at a radius of 100 au in the envelope & 20 K \\
$q$ & Power-law index of the temperature profile in the envelope\tablenotemark{a} & $-0.4$ \\
$r_{\rm d}$ & Radius of the disk\tablenotemark{a} & 20 au \\
$h_{\rm d}$ & Height of the disk\tablenotemark{a} & 10 au \\
$\rho_{\rm disk}$ & Density in the disk & $6.6\times10^{11}$ cm$^{-3}$ \\
$T_{\rm disk}$ & Temperature in the disk & 109 K\\
\hline
& Fitting to the C$^{18}$O (2--1) data & \\
\hline
$v_r$ & Infalling velocity at a radius of 100 au in the envelope  & 1.2 km s$^{-1}$\\
$p_r$ & Power-law index of the velocity profile of the envelope infall  & $-0.6$\\
$v_\phi$ & Rotational velocity at a radius of 100 au in the envelope  & 0.6 km s$^{-1}$\\
$p_\phi$ & Power-law index of the velocity profile of the envelope rotation  & $-1.0$\\
$X_{\rm C^{18}O}$ & C$^{18}$O abundance & $5.5\times10^{-8}$\\
$r_{\rm dep}$ & Transitional radius in the C$^{18}$O abundance profile & 470 au \\
$f_{\rm dep}$ & Depletion factor of the C$^{18}$O abundance outside $r_{\rm dep}$ & 19 \\
$v_{\rm sys}$ & Systemic velocity & 9.1 km s$^{-1}$\\
$M_\star$ & Stellar mass\tablenotemark{a} & 0.08 $M_\sun$ \\
$v_{\rm tur}$ & Turbulent line width\tablenotemark{a} & 0.2 km s$^{-1}$
\enddata 
\tablenotetext{a}{Fixed parameter.}
\tablecomments{The model envelope has power-law profiles of density, temperature, and infalling and rotational velocities and is flattened by $\sin^f\theta$. The density and temperature in the model disk are assumed to be uniform, and its motion is simple Keplerian rotation around a point mass. A gas-to-dust mass ratio of 100 is assumed in the model. The radial profile of the C$^{18}$O abundance in the model is assumed to be a step function.}
\end{deluxetable*}\label{bestfit}

\subsubsection{Infalling and rotational motions}\label{sec_alma_c18o}

\begin{figure*}
\centering
\includegraphics[width=\textwidth]{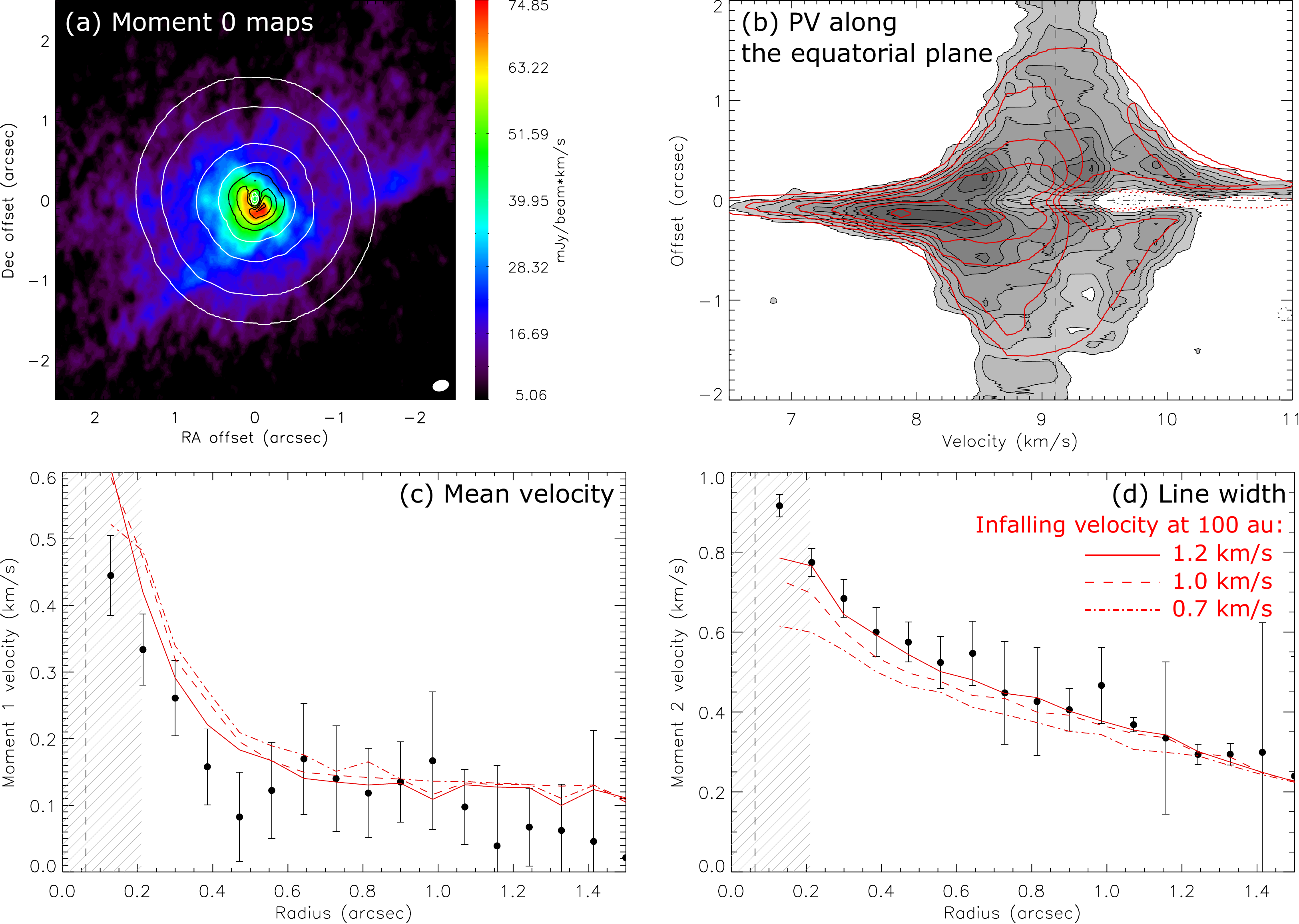}
\caption{(a) Moment 0 map of the C$^{18}$O (2--1) emission observed in HH~211 with ALMA (color scale) overlaid with the synthetic C$^{18}$O (2--1) moment 0 map generated from the visibility data of our best-fit kinematical model (contours). Contours are from 5$\sigma$ in steps of a factor of two, where 1$\sigma$ is 1.7 mJy beam$^{-1}$ km s$^{-1}$. (b) Comparison between the observed (gray scale) and model (red contours) position--velocity diagrams of the C$^{18}$O emission along the midplane of the protostellar envelope. Contours are from 3$\sigma$ in steps of 2$\sigma$. (c) \& (d) Radial profiles of the mean velocity and line width of the C$^{18}$O emission along the midplane observed in HH~211 with ALMA (black data points), where the northeastern and southwestern sides are averaged. Red curves show the profiles measured from the synthetic C$^{18}$O image cubes generated from our kinematical models with different infalling velocities, 1.2 (solid), 1.0 (dashed), 0.7 (dashed-dotted) km s$^{-1}$ at a radius of 100 au. In the model, the infalling velocity is a power-law function with a power-law index of $-0.6$. The infalling velocity of 1.2 km s$^{-1}$ is the best fit, which is shown in (b). Hatched regions and vertical dashed lines denote the beam size and the disk radius, respectively.}\label{alma_c18o}
\end{figure*}

Figure \ref{alma_c18o}a and b present the moment 0 map and position--velocity (PV) diagram along the major axis of the protostellar envelope of the C$^{18}$O emission in HH~211 obtained with the ALMA observations. 
The C$^{18}$O moment 0 map shows a central dip at the protostellar position, 
and in the central region with a diameter of $\sim$1$\arcsec$, the southwestern part is brighter than the northeastern part. 
On a larger scale, the C$^{18}$O emission is extended along the northwest--southeast direction with a length of $\sim$4$\arcsec$ and possibly associated with the outflow in HH~211. 
The PV diagram shows that the southwestern and northeastern parts are blue- and redshifted, respectively, exhibiting a velocity gradient along the major axis of the protostellar envelope. 
In addition, there is negative absorption at redshifted velocities at the center, 
and the blueshifted emission is stronger than the redshifted emission.  
The redshifted absorption and the stronger blueshifted emission around the center suggest that the protostellar envelope is infalling \citep{Evans99}, 
and the velocity gradient along the major axis seen in the PV diagram suggests the rotational motion in the envelope. 
These ALMA C$^{18}$O maps have been discussed in further details in \citet{Lee19}, 
and the velocity channel maps and spectra of the C$^{18}$O emission are presented in Appendix~\ref{app_c18o}.

We have constructed kinematical models of an infalling and rotating protostellar envelope with an embedded Keplerian disk and performed fitting to the C$^{18}$O (2--1) data to measure the infalling and rotational velocities in the envelope around HH~211.
The same density and temperature profiles of the envelope and disk in our best-fit model for the continuum data are adopted in our kinematical model for the C$^{18}$O data. 
The infalling ($v_r$) and rotational ($v_\phi$) velocities are assumed to be power-law functions in spherical coordinates as, 
\begin{eqnarray}
v_r(r) & = & v_{r_0} (\frac{r}{r_0})^{p_r}, \\
v_\phi(r_{\rm rot}) & = & v_{\phi_0} (\frac{r\sin\theta}{r_0})^{p_\phi}, 
\end{eqnarray}
where $r_0$ is set to be 100 au and $\theta$ is the polar angle. 
We note that these assumed velocity profiles in the model envelope are not coupled with the mass distribution in the model or other physical quantities. 
The velocity profiles estimated from our fitting with the kinematical models are compared with the expectation from the mass distribution in HH~211 in the later paragraph.
Considering gas-phase C$^{18}$O can freeze out onto dust grains when the temperature is low at outer radii, 
the C$^{18}$O abundance ($X_{\rm c^{18}o}$) is assumed to be a step function, meaning that it is a constant within a radius $r_{\rm dep}$ and is depleted by a factor of $f_{\rm dep}$ outside $r_{\rm dep}$ as,  
\begin{eqnarray}
X_{\rm c^{18}o}(r) & =  X_{\rm c^{18}o} & {\rm at\ }  r < r_{\rm dep},  \nonumber \\ 
& =  \frac{X_{\rm c^{18}o}}{f_{\rm dep}} & {\rm at\ } r > r_{\rm dep}. 
\end{eqnarray}
The systemic velocity ($v_{\rm sys}$) is adopted as a free parameter.
The turbulent line width ($v_{\rm turb}$) is assumed to be a constant of 0.2 km s$^{-1}$, 
and the protostellar mass ($M_\star$) is fixed and adopted to be 0.08 $M_\odot$, which is estimated from the Keplerian rotation in \citet{Lee18}. 
The motion in the model disk is assumed to be Keplerian rotation around a point mass,  
even though the disk mass could be comparable to the protostellar mass, which could cause disk rotation to deviate from simple Keplerian rotation by 10\% at a radius of 10 au and by 45\% at a radius of 20 au, assuming a disk with a uniform density. 
Since the disk radius is three times smaller than the angular resolution of the C$^{18}$O data, 
we do not expect differences to be distinguished and affect our fitting results of the gas kinematics in the protostellar envelope.
Thus, there are eight free parameters in our kinematical model, which are $v_{r_0}$, ${p_r}$, $v_{\phi_0}$, ${p_\phi}$, $X_{\rm c^{18}o}$, $f_{\rm dep}$, $r_{\rm dep}$, and $v_{\rm sys}$.

The radiative transfer calculations of C$^{18}$O (2--1) were also performed using SPARX to generate model image cubes on the assumption of the LTE,  
and the same fitting procedure was conducted as the fitting to the continuum data. 
The best-fit parameters are listed in Table~\ref{bestfit}, 
and the uncertainties in these parameters due to the noise in the data are less than 3\%.
The corner plots of our fitting to the C$^{18}$O data are presented in Appendix~\ref{app_mcmc}.
Synthetic C$^{18}$O (2--1) images generated from the visibility data of our best-fit model are shown in Fig.~\ref{alma_c18o}. 
Since our model is axisymmetric with power-law density and temperature profiles, 
it cannot explain the larger-scale more irregular morphology of the observed C$^{18}$O emission, 
such as extension along the outflow (northwest--southeast) direction on a 4$\arcsec$ scale. 
Nevertheless, our main focus is the gas kinematics in the inner protostellar envelope within a radius of 1$\arcsec$ (320 au). 
The key observed features within a radius of 1$\arcsec$, including the central dip and asymmetric intensity distribution in the moment 0 map, where the southwestern part is brighter, are also seen in our model moment 0 map. 

Figure~\ref{alma_c18o}b compares the PV diagrams along the midplane of the protostellar envelope extracted from the observed and model images.
Our kinematical model can explain the overall velocity structures in the observed PV diagram, including the blue-skewed line profiles around the center and the velocity of the absorption, although the detailed intensity distribution in the observed PV diagram cannot be reproduced. 
As our primary goal is to measure the gas kinematics in the protostellar envelope, 
we compare the radial profiles of the intensity-weighted mean velocity and line width (i.e., moment 1 and 2) along the midplane between the observations and the models in Fig.~\ref{alma_c18o}c and d.
The error bars in Fig.~\ref{alma_c18o}c and d present the 1$\sigma$ uncertainties in the observed mean velocities and line widths at given radii, which are from the error propagation of the noise in the data.
Because HH~211 is an edge-on source and the infall is faster than the rotation in the envelope, 
the mean velocity and line width along the midplane are primarily related to the rotational and infalling velocities, respectively. 
We subtract the model profiles from the observed profile and compute the residuals. 
For both the velocity and line width, the mean differences between the observed and model profiles are found to be less the 1$\sigma$ uncertainties in these measurements.
Thus, our best-fit model indeed can well explain the observed profiles of the mean velocity and line width along the midplane. 

The uncertainty in the infalling velocity from the model fitting due to the noise is 3\%.
To have a more conservative estimate of the uncertainty in the measured infalling velocity by only considering the velocity features, 
we generated two other model images of the C$^{18}$O emission by reducing the infalling velocity in the kinematical model by 20\% and 40\% compared to the best fit, 
and other parameters were kept the same as the best fit. 
The radial profiles of the mean velocity and line width extracted from these two model images are shown in  Fig.~\ref{alma_c18o}c and d.
The profiles of the mean velocity remain very similar to the best-fit model as expected. 
It is because the mean velocity is more sensitive to the rotational velocity, which was unchanged. 
The profile of the line width at radii less than 1$\arcsec$ starts to deviate from the observations by more than the 1$\sigma$ error bars, when the infalling velocity at a radius of 100 au in the model becomes lower than 1.0 km s$^{-1}$ (i.e., 20\% of the best-fit value).
We have also tested that if the turbulent line width was adopted to be 0.15 or 0.25 km s$^{-1}$, the best-fit infalling velocity would vary by 10\%.
Thus, we expect that the uncertainty in our estimated infalling velocity is better than 20\%.

\begin{figure}
\centering
\includegraphics[width=0.45\textwidth]{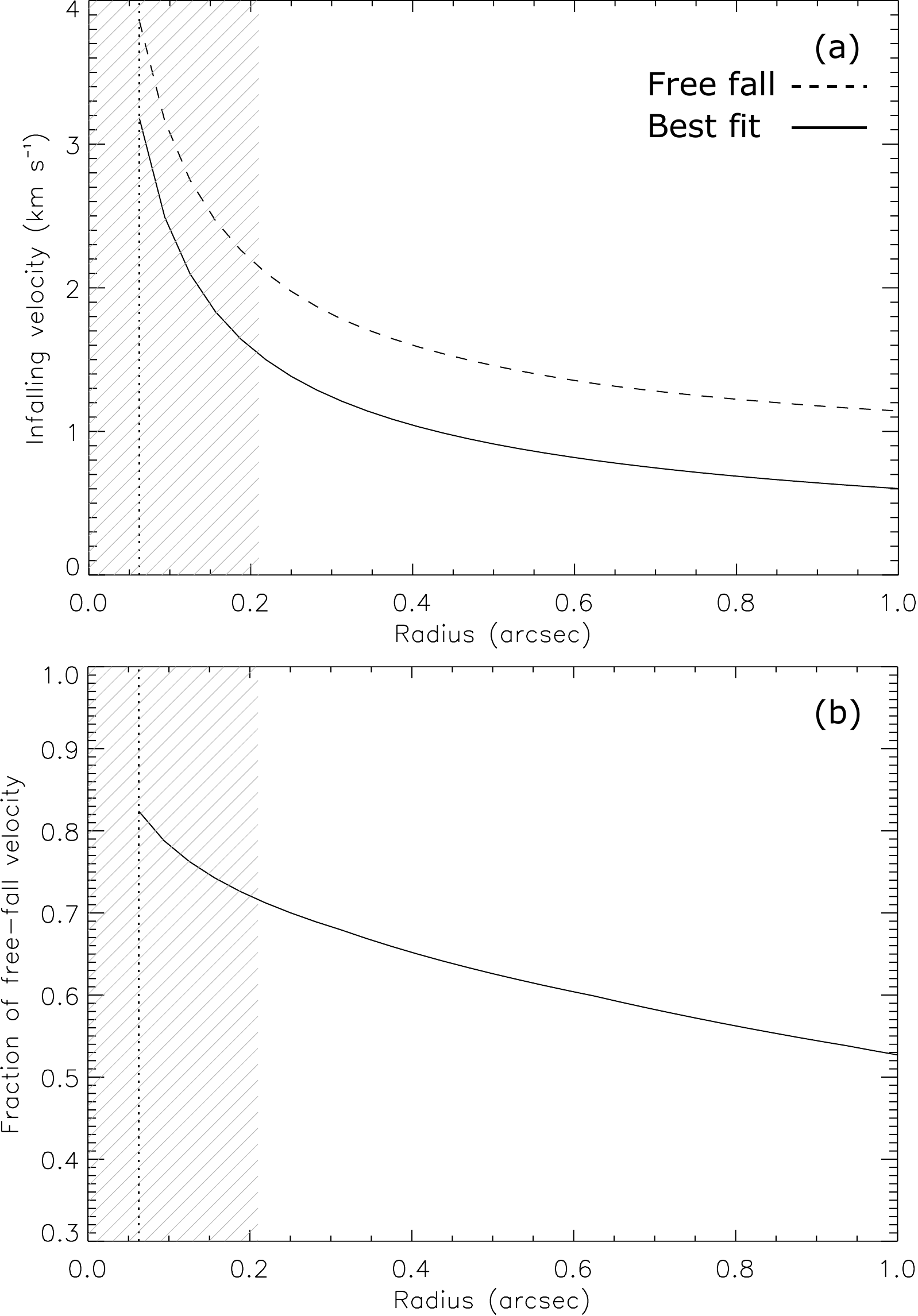}
\caption{(a) Radial profile of the infalling velocity in the best-fit kinematical model in comparison with the radial profile of the free-fall velocity computed from the enclosed mass, including the protostellar, disk, and envelope masses. (b) Radial profile of the ratio of the infalling to free-fall velocities. Hatched regions and vertical dotted lines denote the beam size of the C$^{18}$O data and the disk radius, respectively.}\label{alma_vr}
\end{figure}

In Fig.~\ref{alma_vr}, we compare our estimated infalling velocity from the best-fit kinematical model with the expected free-fall velocity from the mass distribution in HH~211.
The expected free-fall velocity ($v_{\rm ff}$) is calculated from the protostellar mass in the literature \citep{Lee18} and our estimated disk and envelope masses from the continuum data as 
\begin{equation}
v_{\rm ff}(r) = \sqrt{\frac{2GM_{\rm enc}(r)}{r}},
\end{equation}
where $G$ is the gravitational constant and $M_{\rm enc}(r)$ is the enclosed mass within a radius of $r$.
At radii between 0\farcs2 and 1$\arcsec$ (64--320 au), the infalling velocity is estimated to be 50\%--70\% of the free-fall velocity in HH~211. 
The uncertainties in our estimated disk and envelope masses due to the noise in the data are small ($<$3\%).
However, if the disk and envelope masses are actually lower due to larger dust absorption coefficients at 0.8 mm and 1.3 mm, 
the ratio of the infalling to free-fall velocities would become higher.

\subsubsection{Magnetic field structure and curvature}

\begin{figure}
\centering
\includegraphics[width=0.42\textwidth]{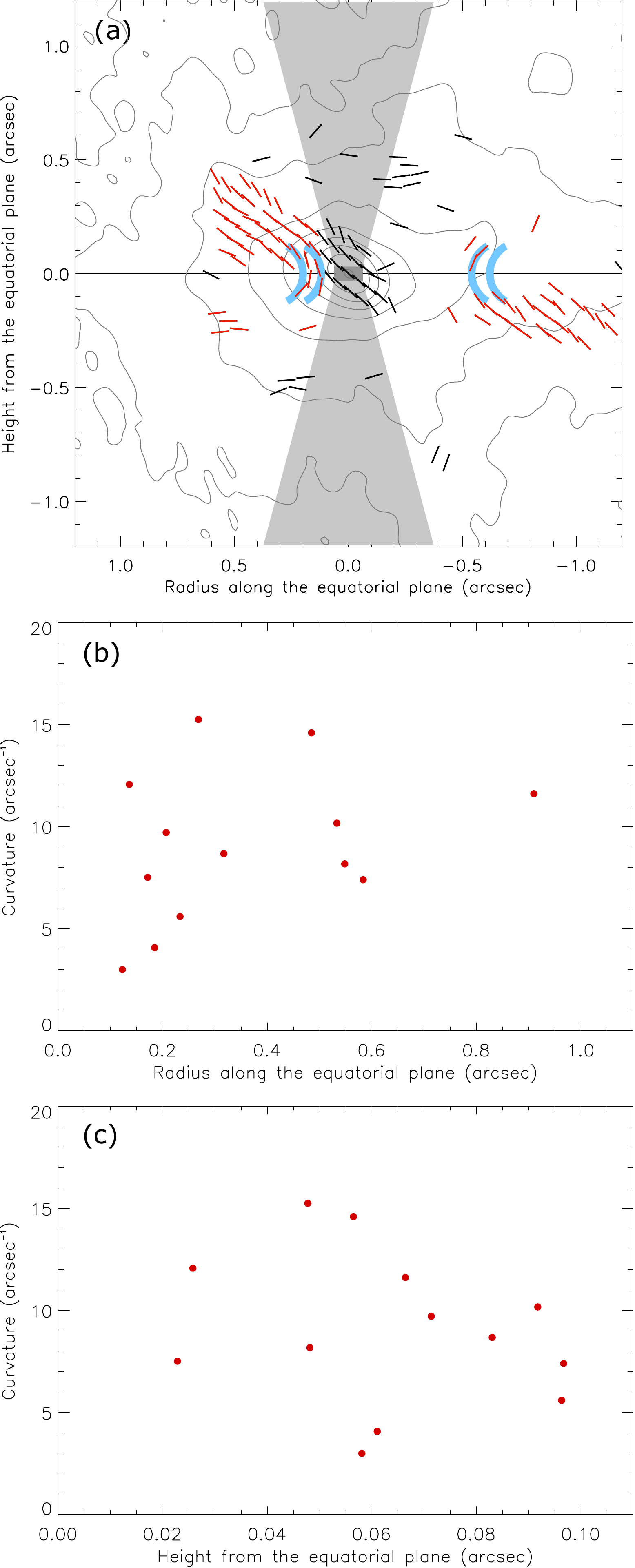}
\caption{(a) Magnetic field (black and red segments) observed in HH~211 with ALMA. Contours show the 0.8~mm continuum emission and are from 5$\sigma$ in steps of a factor of two. The map is rotated to have the midplane along the horizontal direction. Dark and light gray shaded areas present the disk and outflow components, respectively. Red segments likely trace the pinched magnetic fields in the protostellar envelope. Blue arcs have a curvature of 7.1 arcsecond$^{-1}$, corresponding to a curvature radius of 0\farcs14, for comparison with the observed magnetic field segments. Black segments are located close to the optically-thick disk and the outflow, so they may not be related to the magnetic fields in the protostellar envelope. (b) \& (c) Estimated curvature of the magnetic fields as functions of radius along the midplane and vertical height from the midplane using the red segments within 0\farcs1 from the midplane.}\label{alma_pol}
\end{figure}

Figure~\ref{alma_pol}a presents the magnetic field structures in the protostellar envelope around HH~211 observed with ALMA, 
which have also been discussed in detail in \citet{Lee19}.
The map is rotated to have the midplane of the envelope along the horizontal axis. 
HH~211 is almost edge-on with an inclination angle of 81$\arcdeg$. 
The toroidal magnetic fields are expected to be canceled out along the line of sight. 
The map primarily shows the pinched poloidal magnetic fields, as shown by red segments in Fig.~\ref{alma_pol}a. 
The polarization detections close to the optically-thick disk are possibly caused by dust scattering \citep{Cox18}, 
while those close to the outflows can be affected by the outflow activities \citep{Hull17}. 
These polarization detections, shown by black segments, may not be related to the magnetic fields in the protostellar envelope and are excluded in our analysis.

To assess the balance between the magnetic field tension force and the gravity in the protostellar envelope on a 600 au scale in HH~211, 
we first estimated the curvature of the magnetic fields in the midplane. 
The ALMA polarization data suggests that the magnetic fields in the protostellar envelope in HH~211 have been dragged inward along the midplane. 
As shown in theoretical simulations of magnetic fields in infalling protostellar envelopes \citep{Mellon08, Zhao18}, 
the curvature of the magnetic field increases as the distance to the midplane decreases, 
and the direction of the curvature of the magnetic fields close to the midplane is expected to be along the midplane. 
Thus, we selected the magnetic field segments with their distances to the midplane less than 0\farcs1, approximately half the beam size, 
and we assumed that the center of the curvature of the magnetic fields traced by the selected segments is in the midplane.
Then, the curvature of the magnetic field traced by each segment can be measured as the inverse of the distance between the segment and the midplane along the direction perpendicular to its magnetic field orientation. 

Fig.~\ref{alma_pol}b and c present the measured curvatures from the selected segments as functions of radius along the midplane and vertical height from the midplane, respectively.
The measured curvature of the magnetic fields close to the midplane ranges from 3 to 16 arcsecond$^{-1}$, corresponding to a mean curvature radius of 0\farcs14 with a standard deviation of 0\farcs08, 
and there is no significant variation in the curvature from outer to inner radii.
In Fig.~\ref{alma_pol}a, thick light blue arcs having a curvature radius of 0\farcs14 are plotted as an example for comparison with the observed magnetic field orientations.  
We have also estimated the curvature of the magnetic fields from the separation and difference in their orientations between two neighboring segments, as Equation~6 in \citet{Koch12}. 
With several pairs of the magnetic field segments close to the midplane, delineating bent magnetic fields as those around the blue arcs in Fig.~\ref{alma_pol}a, the mean curvature of the magnetic fields is estimated to be 5 arcsecond$^{-1}$, corresponding to a curvature radius of 0\farcs2, consistent with the estimates in Fig.~\ref{alma_pol}b and c. 

\subsubsection{Magnetic field strength}\label{sec_env_b}

The strength of the poloidal magnetic field ($B$) in an infalling protostellar envelope can be estimated from the balance between the infalling motion, gravity, and magnetic field tension along the radial direction in the midplane on the assumption of a steady state as,
\begin{equation}\label{force}
\rho(r)v_r(r)\frac{dv_r}{dr}\approx-\frac{GM_{\rm enc}(r)\rho(r)}{r^2}+\frac{{B(r)}^2}{4\pi R_{\rm cur}(r)},
\end{equation}
where $G$ is the gravitational constant, $M_{\rm enc}$ is the enclosed mass within a radius of $r$, and $R_{\rm cur}$ is the curvature radius of the poloidal magnetic field, as discussed in the previous studies \citep{Koch12,Aso15,Sai22}. 
The magnetic pressure is not considered here because it is expected to be negligible compared to the magnetic tension and gravity \citep{Zhao18}, 
and the observations do not show tangled magnetic fields on the scale of the protostellar envelope (Fig.~\ref{alma_pol}). 
We have also estimated the force caused by the thermal pressure gradient in the protostellar envelope to be approximately 10\% of the gravitational force, based on our best-fit model to the continuum data. The contribution by the thermal pressure is less than our uncertainty in the infalling velocity. 
Thus, the thermal pressure is also not considered here.
We assume that infalling velocity is a fraction of the free-fall velocity ($v_{\rm ff}$), as $v_r(r) = \alpha v_{\rm ff}(r)$.
Then, from Equation~\ref{force}, the magnetic field strength at radius $r$ can be expressed as, 
\begin{equation}\label{G-Btension}
B=\sqrt{(1-\alpha^2)4\pi R_{\rm cur}\frac{GM_{\rm enc}\rho}{r^2}},  
\end{equation}
where $\alpha$ is the ratio of the infalling to free-fall velocities \citep{Sai22}.
Here we assume that $\alpha$ is a constant for simplicity, since our data may not be able to resolve the actual radial profile of the infalling velocity, which could be complicated as seen in theoretical simulations and may not be fully described with a power-law function \citep{Mellon08, Zhao18}. 
Further studies to apply this method on synthetic images from theoretical simulations can help with quantifying any bias and calibration of Equation~\ref{G-Btension}.

We estimated the mean magnetic field strength in the protostellar envelope within a radius of 1$\arcsec$ (320 au). 
The envelope mass within a radius of 1$\arcsec$ was estimated to be 0.06 $M_\sun$ from our model fitting to the continuum emission, 
and thus the mean density $\rho$ in the inner envelope and the enclosed mass $M_{\rm enc}$ within a radius of 1$\arcsec$ were estimated to be 2.8$\times$10$^{-16}$ g cm$^{-3}$ and 0.24 $M_\sun$, respectively, where the protostellar and disk masses are 0.08 and 0.09 $M_\sun$.
We adopted $\alpha$ to be 0.6, a median ratio of the infalling to free-fall velocities within a radius of 1$\arcsec$ (Fig.~\ref{alma_vr}), 
and the curvature radius $R_{\rm cur}$ was adopted to 0\farcs14.
Then the magnetic field strength in the protostellar envelope on a 600 scale was estimated to be 0.7 mG. 
We computed the mean column density as $M_{\rm enc}/\pi r^2$ and estimated the mass-to-flux ratio in the envelope to be 14.8 with Eq.~\ref{mtf}.

We note that if a larger dust absorption coefficient is adopted in the estimate, 
which results in a lower $M_{\rm enc}$ and a larger $\alpha$ and means that the infall is actual close to the free fall, 
the estimated magnetic field strength would become smaller, 
and the estimated mass-to-flux ratio would become larger. 
If the disk and envelope masses are 30\% lower, the estimated magnetic field strength and mass-to-flux ratio are 35\% smaller and 20\% larger, respectively. 
On the other hand, 
if the infalling velocity is over-estimated by 20\%, the estimated magnetic field strength would become 30\% larger.  
In addition, we note that we only considered the poloidal component of the magnetic field. 
The presence of the toroidal magnetic field has been suggested in \citet{Lee18} because of the asymmetric distribution of the polarized intensity, 
which could be due to wrapped magnetic field lines. 
Nevertheless, we expect that the toroidal magnetic field is weaker than the poloidal magnetic field because the rotation is twice slower than the infall in the protostellar envelope in HH~211.
Therefore, considering the uncertainties in the infalling velocity and curvature radius and inclusion or exclusion of the toroidal component of the magnetic field, which is at most as strong as the poloidal magnetic field, 
the magnetic field strength and mass-flux-ratio in the protostellar envelope on a 600 au scale around HH~211 are estimated to be in the ranges of 0.3--1.2 mG and 9.1--32.3, respectively.
Similar to our estimates of the magnetic field strength on the core scale, 
these values are subject to the adopted dust absorption coefficient. 

\section{Discussion}\label{discussion}

\subsection{Magnetic field strength in the protostellar envelope and the DCF method}\label{discussion1}
\begin{figure}
\centering
\includegraphics[width=0.45\textwidth]{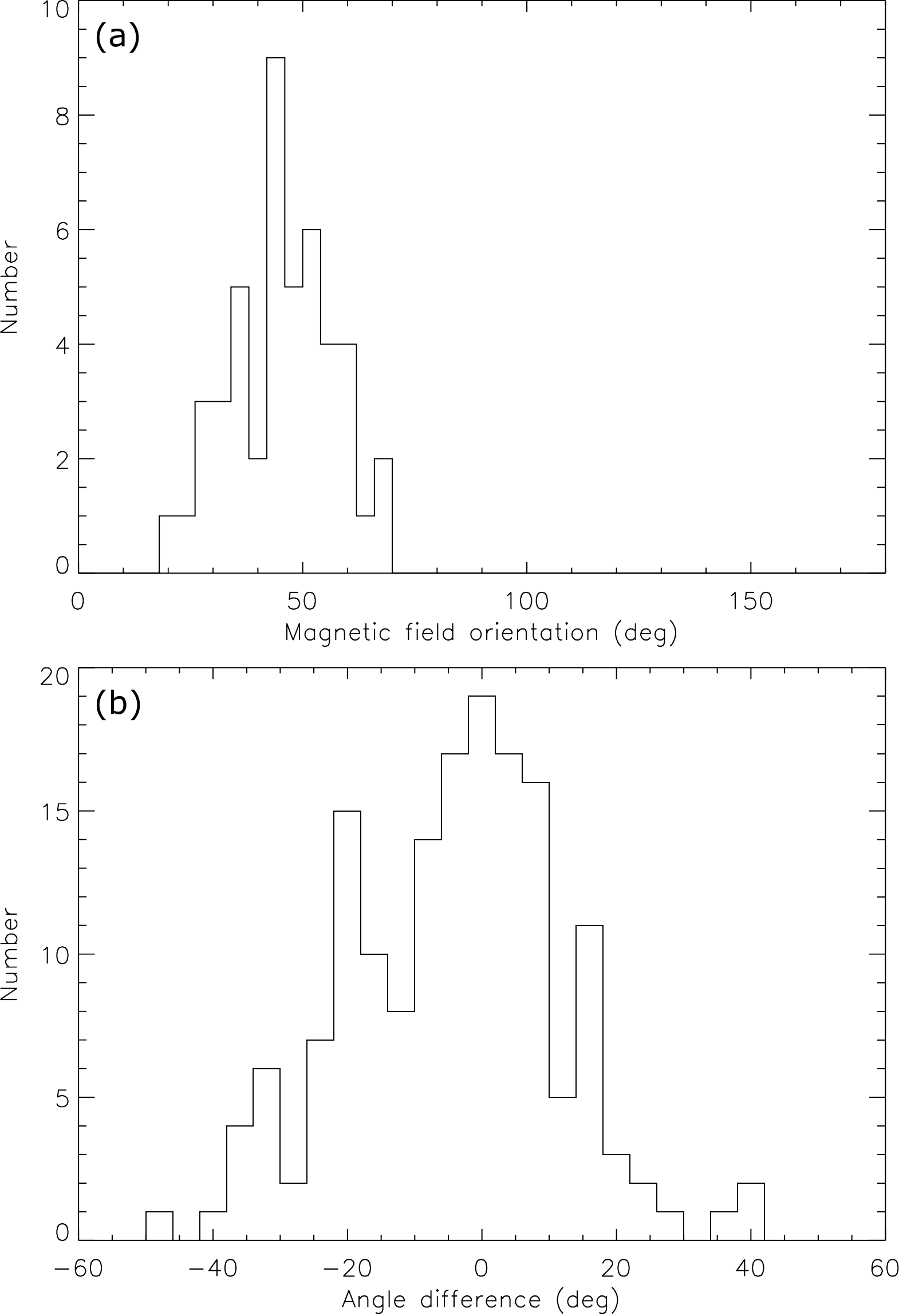}
\caption{(a) Histogram of the orientations of the magnetic field segments in the second and forth quadrants and at radii larger than 0\farcs3 or smaller than $-0\farcs7$ along the midplane in Fig.~\ref{alma_pol}. The orientations are defined as angles from the normal axis of the protostellar envelope counterclockwise, meaning that a segment with an angle of 90$\arcdeg$ is in the direction parallel to the midplane of the envelope. (b) Histogram of the  angle differences between pairs of the segments with the separation of one beam size in (a).}\label{alma_dcf}
\end{figure}

The magnetic field strength in the protostellar envelope on a 600 au scale in HH~211 is estimated to be 0.3--1.2 mG, from the balance between the infalling motion, gravity, and magnetic field tension.
For comparison, we have also applied the DCF method to the ALMA polarization data to estimate the magnetic field strength in the protostellar envelope.
Figure~\ref{alma_pol} shows that the magnetic field structures change significantly within a scale comparable to the beam size of 0\farcs18, especially in the regions close to the midplane, and the spatial coverage of the polarization detections is sparse. 
Thus, it is not straightforward to remove the overall magnetic field structures and estimate the angular dispersion of the magnetic field orientations, 
so we only selected the magnetic field segments in the second and forth quadrants and at radii larger than 0\farcs3 or smaller than $-0\farcs7$ along the midplane in Fig.~\ref{alma_pol}, where the magnetic field structures are more uniform, to estimate the angular dispersion. 
Figure~\ref{alma_dcf}a presents the distribution of the orientations of the selected magnetic field segments, 
and the 1$\sigma$ width of the distribution is estimated to be 12$\arcdeg$. 
We computed the angle difference between each pair of the segments with the separation of one beam size. 
Figure~\ref{alma_dcf}b presents a histogram of the angle differences. 
The standard deviation of these angle differences was computed to be 16$\arcdeg$ and adopted as the observed angular dispersion. 
The typical uncertainty in the magnetic field orientations inferred from the ALMA polarization data is 8$\arcdeg$. 
Thus, the intrinsic angular dispersion of the magnetic field orientations in the protostellar envelope, excluding the scatter due to the noise, was estimated to be $\sqrt{16\arcdeg^2-8\arcdeg^2}=14\arcdeg$. 
We assumed that the turbulent velocity in the protostellar envelope on a 600 au scale in HH~211 is the same as the thermal sound speed of 0.2 km s$^{-1}$ at 10~K because the temperature was estimated to be 10--20 K at radii between 500 and 100 au with our analysis of the ALMA continuum data (Section \ref{sec_alma_cont}).  
With the angular dispersion of 14$\arcdeg$, the mean density of 2.8$\times$10$^{-16}$ g cm$^{-3}$ in the envelope and Equation~\ref{dcf}, the magnetic field strength in the protostellar envelope on a 600 au scale was estimated to be 5 mG with the DCF method.

The magnetic field strength estimated with the DCF method is significantly higher than our estimate considering the force balance based on the infalling velocity, mass distribution, and magnetic field curvature in the protostellar envelope.
The magnetic fields in the protostellar envelope in HH~211 are highly pinched with a mean curvature of 5--7 arcsecond$^{-1}$ in the midplane (Fig.~\ref{alma_pol}). 
We found that if the strength of the poloidal magnetic field in the protostellar envelope is larger than 2 mG, 
the magnetic field tension force would be larger than the gravitational force in the protostellar envelope (Eq.~\ref{force}). 
On the other hand, 
the infalling motion has been observed in the protostellar envelope in HH~211, 
and the infalling velocity is higher at a smaller radius (Section \ref{sec_alma_c18o}), 
so the gravity is expected to dominate over the magnetic field tension. 
This suggests that the DCF method likely overestimates the magnetic field strength on the envelope scale in HH~211. 
This is possibly because in a dynamically infalling protostellar envelope, the magnetic fields are being dragged inward and the field structures could change significantly even over a small scale, as seen in theoretical simulations \citep{Mellon08, Zhao18}, 
and the observed angle differences between the nearby magnetic field segments may not be related to the turbulence. 
Furthermore, in a dynamically infalling protostellar envelope, it could be degenerate to separate the velocity component of the turbulence from the infalling and rotational motions in the observed velocity pattern.
Thus, the assumptions of the DCF method are probably not valid in this case.

\subsection{Change in the magnetic flux from large to small scales}
The magnetic field strength in the dense core on a 0.1 pc scale in HH~211 is estimated to be 40--107 $\mu$G with the DCF method. 
The estimated magnetic field strength on a 0.1 pc scale in HH~211 is comparable to those in many other dense cores with similar densities estimated with the DCF method and Zeeman observations \citep{Crutcher12,Tritsis15,Myers21}.
The dense core associated with HH~211 has a size of $80\arcsec \times 40\arcsec$ with a position angle of the major axis of 48$\arcdeg$ and has a mean density of (3.3--6.6)$\times$10$^{-19}$ g cm$^{-3}$ or (1--2)$\times$10$^5$ cm$^{-3}$.
On a smaller scale of 600 au, the protostellar envelope approximately has a size of $2\farcs5 \times 1\farcs3$ with a position angle of the major axis of 36$\arcdeg$, as observed in the 1.3 mm and 0.8 mm continuum with ALMA \citep{Lee19}, and has a mean density of 2.8$\times$10$^{-16}$ g cm$^{-3}$ or 8$\times$10$^7$ cm$^{-3}$.
The shape and orientation of the protostellar envelope on the small scale are similar to those of the dense core on the large scale.
Theoretical calculations show that when a spheroid contracts with its shape unchanged under the flux-freezing condition, 
the magnetic field strength would scale as $\rho^{2/3}$ \citep{Myers18}.
Therefore, in HH~211, if the magnetic flux is frozen in the matter in the dense core, 
the magnetic field strength in the protostellar envelope is expected to be 3.4$\pm$1 mG.

As discussed in Section \ref{discussion1}, the magnetic field strength in the protostellar envelope in HH~211 is expected to be smaller than 2 mG and is likely in the range of 0.3--1.2 mG based on the observed gas kinematics and magnetic field structures, and this estimated magnetic field strength on the envelope scale includes both poloidal and toroidal components.
The estimated magnetic field strength in the protostellar envelope is lower than the expectation from the flux-freezing condition, and
our results suggest that the magnetic field strength increases from the scales of the dense core to the protostellar envelope with a scaling relation of $B \propto \rho^{0.36\pm0.08}$.
Furthermore, we note that the magnetic field strength estimated on the dense core scale only include the plane-of-sky component and does not include correction for the unknown inclination of the magnetic field, 
so the magnetic field strength could be actually larger and the scaling relation could be even shallower. 
The estimated magnetic field strengths and the parameters adopted for the estimates on the large and small scales are summarized in Table \ref{bfield}.

Therefore, our results show that the scaling relation between the magnetic field strength and density in HH~211 is shallower than the theoretical expectation $B \propto \rho^{1/2}$ or $\rho^{2/3}$ for oblate or spherical contraction of dense cores under the flux-freezing condition \citep{Tritsis15,Hennebelle19}. 
A shallow scaling relation, like $B \propto \rho^{0}$, could occur if the structures only contract along the magnetic fields. 
This scenario may not fully explain the case in HH~211 because the infalling motion along the midplane, which is across the magnetic fields, has been observed. 
Therefore, our results may hint at the other possibility of the magnetic field being only partially coupled with the matter on 0.1 pc to 600 au scales in the dense core in HH~211.

\begin{deluxetable*}{ccccccc}
\tablecaption{Physical conditions and magnetic field strengths in HH~211}
\centering
\tablehead{\multicolumn{7}{c}{Dense core on a 0.1 pc scale}}
\startdata
$\rho$ (g cm$^{-3}$) & $\delta v_{\rm nt}$ (km s$^{-1}$) & $\delta \theta_{\rm B}$ (\arcdeg) & $\Sigma$ (g cm$^{-2}$)  & $B$ ($\mu$G) & $\lambda$ & \\
\hline 
(3.3--6.6)$\times$10$^{-19}$ & 0.15 & 12--15 & 0.09 & 40--107 & 1.2--3.7 & \\
\hline \hline
\multicolumn{7}{c}{Protostellar envelope on a 600 au scale} \\
\hline
$\rho$ (g cm$^{-3}$) & $M_{\rm enc}$ ($M_\sun$) & $R_{\rm cur}$ & $\alpha$ & $\Sigma$ (g cm$^{-2}$) & $B$ (mG) & $\lambda$ \\
\hline
2.8$\times$10$^{-16}$ & 0.24 & 0\farcs14$\pm$0\farcs08 & 0.6 & 6.6 & 0.3--1.2 & 9.1--32.3
\enddata 
\tablecomments{This table summarizes the parameters adopted to estimate the magnetic field strengths in the dense core on a 0.1 pc scale with the DCF method using Eq.~\ref{dcf}--\ref{mtf} and in the protostellar envelope on a 600 au scale from the force balance between the gravity and magnetic field tension using Eq.~\ref{G-Btension} in HH~211. $\rho$ is the density. $\delta v_{\rm nt}$ is the non-thermal line width. $\delta \theta_{\rm B}$ is the angular dispersion of the magnetic field orientations. $\Sigma$ is the column density. $B$ is the magnetic field strength. $\lambda$ is the mass-to-flux ratio. $M_{\rm enc}$ is the enclosed mass, including the protostellar, disk, and envelope masses, within a radius of 1$\arcsec$ (320 au). $R_{\rm cur}$ is the mean curvature radius of the magnetic fields in the envelope. $\alpha$ is the ratio between the estimated infalling and free-fall velocities.}
\end{deluxetable*}\label{bfield}

\subsection{Mass-to-flux ratios and hints of ambipolar diffusion}
The density ranges from 3.3$\times$10$^{-19}$--2.8$\times$10$^{-16}$ g cm$^{-3}$ (or 1$\times$10$^5$--8$\times$10$^7$ cm$^{-3}$) on scales of 0.1 pc to 600 au in HH~211. 
In this density range, ambipolar diffusion is expected to be the most efficient non-ideal MHD effect \citep{Zhao16,Dzyurkevich17,Tsukamoto20}.
Our results suggest that the mass-to-flux ratio increases by a factor of 2.5 or more from 1.2--3.7 in the dense core on a 0.1 pc scale to 9.1--32.3 in the protostellar envelope on a 600 au scale.
A similar increase in the mass-to-flux ratio on a scale of a few hundred au in collapsing dense cores is seen in the non-ideal MHD simulations with efficient ambipolar diffusion \citep{Zhao18}, in which dust grains with sizes smaller than 0.1 $\mu$m were excluded to enhance the diffusivity of ambipolar diffusion by one to two orders of magnitude compared to that computed with the standard MRN dust size distribution \citep[Mathis-Rumpl-Nordsieck;][]{Mathis77}.
Such enhanced ambipolar diffusion is also seen in numerical simulations incorporating dust coagulation with the standard MRN dust size distribution \citep{Guillet20}.
On the other hand, in other non-ideal MHD simulations with magnetic diffusivities computed with the standard MRN dust size distribution, 
a significant increase in the mass-to-flux ratio only occurs on a scale smaller than 100 au \citep{Masson16,Zhao18}. 

Although the mass-to-flux ratio can increase even in the ideal MHD limit when there is turbulence, 
the magnetic diffusion due to the turbulence is expected to become significant on a small scale of 100 au, as seen in the numerical simulations \citep{Joos13,G-C16}.
The mass-to-flux ratio on a larger scale of a few hundreds au may increase by a factor of a few in the ideal MHD simulation with supersonic turbulence \citep{S-L12,S-L13,Joos13,Seifried13}, 
but the turbulent velocity in the dense core in HH~211 is measured to be subsonic (Section~\ref{sec_45m_n2hp}). 
The mass-to-flux ratio also can increase due to expansion of magnetic structures formed by the magnetic flux decouples from the material accreted onto a protostar, the so-called ``decoupling-enabled magnetic structure'' \citep[DEMS;][]{Zhao11,Lam19}.
Nevertheless, the presence of DEMS tends to disrupt disk formation, and no persistent disk forms in the numerical simulations with strong DEMS \citep{Zhao18,Lam19}, 
while the Keplerian disk with a radius of 20 au has been observed in HH~211. 
These mechanisms are less likely to explain the increase in the mass-to-flux ratio and the shallow $B\mbox{--}\rho$ relation from the dense core to protostellar envelope scales in HH~211. 

Therefore, 
our results of the increase in the mass-to-flux ratio from the large to small scales could hint at efficient ambipolar diffusion in the protostellar envelope on a scale of several hundred au in HH~211, which enables decoupling between the magnetic field and the neutral matter. 
As a result, less magnetic flux can be transported to the protostellar envelope, and the efficiency of magnetic braking can be reduced. 
Other non-ideal MHD effects may be still important on a smaller scale with a higher density in HH~211, 
but the magnetic field structures and strengths on a small scale around the Keplerian disk cannot be probed with the resolution of our data.
Nevertheless, in theoretical simulations with ambipolar diffusion and without other non-ideal MHD effects \citep{Masson16, Zhao16, Zhao18}, 
sizable Keplerian disks can form if ambipolar diffusion is efficient to increase the mass-to-flux ratio by a factor of a few from the dense core to protostellar envelope scale, as the case in HH~211.
In \citet{Masson16}, the disk grow to have an outer radius of 15--30 au when the total mass of the central object and disk reaches 0.2 $M_\sun$. 
In \citet{Zhao18}, the disk radius is $\sim$20 au when the mass of the protostar+disk system is 0.1 $M_\sun$. 
HH~211 has a disk radius of 20 au and a protostar+disk mass of 0.17 $M_\sun$, comparable to the values in these non-ideal MHD simulations with efficient ambipolar diffusion.
Thus, our results could support the scenario of efficient ambipolar diffusion enabling the formation of the 20 au Keplerian disk in HH~211. 
The disk radius in HH~211 is also comparable to the theoretical expectation from the disk formation which is self regulated by the magnetic braking and ambipolar diffusion \citep{Hennebelle16}.  
 In addition, the high-resolution numerical simulations show that the disk becomes less magnetized compared to its surrounding protostellar envelope and the disk properties weakly depend on the physical conditions of the parental dense core, because of the decoupling between the magnetic fields and matter in the inner dense region \citep{Hennebelle20,Lee21}. 
Our results hinting at efficient ambipolar diffusion could also support the importance of the non-ideal MHD effects in the star formation process suggested in these theoretical studies. 
Future studies of magnetic structures and strengths, ionization rates, and dust properties on several different scales in protostellar sources are essential to understand the mechanisms determining the efficiency of ambipolar diffusion.

\subsection{Uncertainty due to the dust absorption coefficients}\label{sec_dkappa}
We note that our estimated field strengths in the dense core and the protostellar envelope are subject to the adopted dust absorption coefficient at millimeter wavelengths, which was assumed to be the same on both large and small scales. 
Nevertheless, the ratio of the magnetic field strengths in the dense core and the protostellar envelope would be valid if the dust absorption coefficient does not significantly change from the large to small scales. 
However, grain growth could occur in dense cores and protostellar disks if the density is high \citep{Ormel09, Bate22}.
Dust emissivity indices $\beta$ smaller than 1 indeed have been observed in a few protostellar sources on a scale of a few hundred au, suggestive of grain growth and presences of millimeter sized grains \citep{Galametz19}.
If the dust absorption coefficient at millimeter wavelengths actually increases on the small scale and becomes similar to those in protoplanetary disks due to grain growth \citep{Beckwith90,Birnstiel18}, 
the estimated magnetic field strength in the protostellar envelope would become smaller, and the mass-to-flux ratio would become larger (Section~\ref{sec_env_b}). 
This would lead to an even shallower scaling relation than $B \propto \rho^{0.36\pm0.08}$ and a larger increase in the mass-to-flux ratios from the scales of the dense core to the inner protostellar envelope.
Therefore, our discussions and conclusions are not affected by possible grain growths in the inner high-density region in HH~211.

\section{Summary}
To study transportation of magnetic flux from large to small scales in protostellar sources, we analyzed the Nobeyama 45-m N$_2$H$^+$ (1--0), JCMT 850~$\mu$m polarization, and ALMA C$^{18}$O (2--1) and 1.3~mm and 0.8~mm (polarized) continuum data of the Class 0 protostar HH~211.

We identified the dense core associated with HH~211 in the N$_2$H$^+$ (1--0) emission using the {\it dendrogram} algorithm, 
and the size of the dense core was estimated to be $80\arcsec \times 40\arcsec$ (0.12 pc $\times$ 0.06 pc). 
The line width in the dense core was measured by fitting the N$_2$H$^+$ (1--0) hyperfine lines, and the turbulent line width was estimated to be 0.15 km s$^{-1}$. 
The mass and density of the dense core were estimated to be 2.5 $M_\sun$ and (3.3--6.6)$\times$10$^{-19}$ g cm$^{-3}$ with the JCMT 850 $\mu$m continuum data, respectively. 
The angular dispersion of the magnetic field orientations in the dense core was measured to be 12$\arcdeg$--15$\arcdeg$ from the JCMT 850 $\mu$m polarization data, after removing the large-scale magnetic field structures. 
With the angular dispersion of the magnetic fields, turbulent line width, and density in the dense core measured from the single-dish data, 
the magnetic field strength on a 0.1 pc scale around HH~211 was estimated to be 40--107 $\mu$G with the DCF method, corresponding to a mass-to-flux ratio of 1.2--3.7. 
This field strength is comparable to those in other dense cores with similar density estimated with Zeeman observations. 

We measured the density and temperature profiles in the protostellar envelope on a 600 au scale around HH~211 by fitting the ALMA 1.3 mm and 0.8 mm continuum visibility data with models of a protostellar envelope with an embedded disk. 
We adopted these density and temperature profiles and constructed kinematical models of an infalling and rotating protostellar envelope to fit the ALMA C$^{18}$O (2--1) visibility data and estimate the infalling velocity. 
We found that the infalling velocity in the protostellar envelope is approximately 60\% of the expected free-fall velocity.
The pinched magnetic fields in the protostellar envelope on a 600 au scale have been revealed with the ALMA 0.8 mm polarization data. 
We measured the mean curvature of the pinched magnetic fields in the envelope midplane to be 5--7 arcsecond$^{-1}$ from the ALMA polarization data. 
With the infalling velocity, curvature of the magnetic fields, and mass distribution in the protostellar envelope, 
we analyzed the force balance between the gravity and magnetic field tension, 
and the magnetic field strength on a 600 au scale around HH~211 was estimated to be 0.3--1.2 mG, corresponding to a mass-to-flux ratio of 9.1--32.3. 
We also found that the application of the DCF method to the ALMA polarization data most likely overestimates the magnetic field strength, which would result in a magnetic field tension force larger than the gravitational force. 

Our analysis suggests a scaling relation between the magnetic field strength and density of $B \propto \rho^{0.36\pm0.08}$ and an increase in the magnetic field by more than a factor of two between 0.1 pc and 600 au scales around HH~211. 
This trend is different from the theoretical expectation from the ideal MHD limit, where the magnetic fields and matter are coupled well.   
Although our estimated magnetic field strengths are subject to the adopted dust absorption coefficients, 
the ratio in the magnetic field strengths remains unchanged if the dust absorption coefficients do not change significantly from the large to small scale. 
Furthermore, this trend is still valid if the dust grain on the small scale evolve to have absorption coefficients similar to those in protoplanetary disks. 
Therefore, our results could hint that the magnetic fields are partially decoupled from the neutral matter in the collapsing dense core in HH~211 due to the non-ideal MHD effects. 
Other mechanisms, such as contraction only along the magnetic field lines, turbulence, and decoupling-enabled magnetic structure, are less likely to explain the increase in the mass-to-flux ratio and the shallow $B-\rho$ relation between the scales of 0.1 pc and 600 au in HH~211. 
In the density range on scales from 0.1 pc to 600 au in HH~211, ambipolar diffusion is the most dominant non-ideal MHD effect. 
A similar increase in the mass-to-flux ratio on a scale of a few hundred au in a collapsing dense core has also be seen in theoretical simulations with efficient ambipolar diffusion. 
Thus, our results could support the scenario of efficient ambipolar diffusion enabling the formation of the 20 au Keplerian disk in~HH 211.

\acknowledgements 
{
We thank Che-Yu Chen for fruitful discussions and suggestions on this project. 
The Nobeyama 45-m radio telescope is operated by Nobeyama Radio Observatory, a branch of National Astronomical Observatory of Japan.
The James Clerk Maxwell Telescope is operated by the East Asian Observatory on behalf of The National Astronomical Observatory of Japan; Academia Sinica Institute of Astronomy and Astrophysics; the Korea Astronomy and Space Science Institute; the National Astronomical Research Institute of Thailand; Center for Astronomical Mega-Science (as well as the National Key R\&D Program of China with No. 2017YFA0402700). Additional funding support is provided by the Science and Technology Facilities Council of the United Kingdom and participating universities and organizations in the United Kingdom and Canada. 
Additional funds for the construction of SCUBA-2 were provided by the Canada Foundation for Innovation.
The authors wish to recognize and acknowledge the very significant cultural role and reverence that the summit of Maunakea has always had within the indigenous Hawaiian community.  We are most fortunate to have the opportunity to conduct observations from this mountain.
This paper makes use of the following ALMA data: ADS/JAO.ALMA\#2016.1.00017.S and 2017.1.01078.S. ALMA is a partnership of ESO (representing its member states), NSF (USA) and NINS (Japan), together with NRC (Canada), NSTC and ASIAA (Taiwan), and KASI (Republic of Korea), in cooperation with the Republic of Chile. The Joint ALMA Observatory is operated by ESO, AUI/NRAO and NAOJ. 
H.-W.Y.\ acknowledges support from the National Science and Technology Council (NSTC) in Taiwan through the grant NSTC 110-2628-M-001-003-MY3 and from the Academia Sinica Career Development Award (AS-CDA-111-M03).
PMK is supported by the National Science and Technology Council in Taiwan through grants NSTC 109-2112-M-001-022, NSTC 110-2112-M-001-057, and NSTC
111-2112-M-001-003.
S.T.\ is supported by JSPS KAKENHI Grant Numbers JP21H00048 and JP21H04495. 
K.T.\ was supported by JSPS KAKENHI (Grant Number 20H05645). 
}

\begin{appendix}
\section{ALMA velocity channel maps and spectra of the C$^{18}$O (2--1) emission}\label{app_c18o}
\begin{figure*}
\centering
\includegraphics[width=\textwidth]{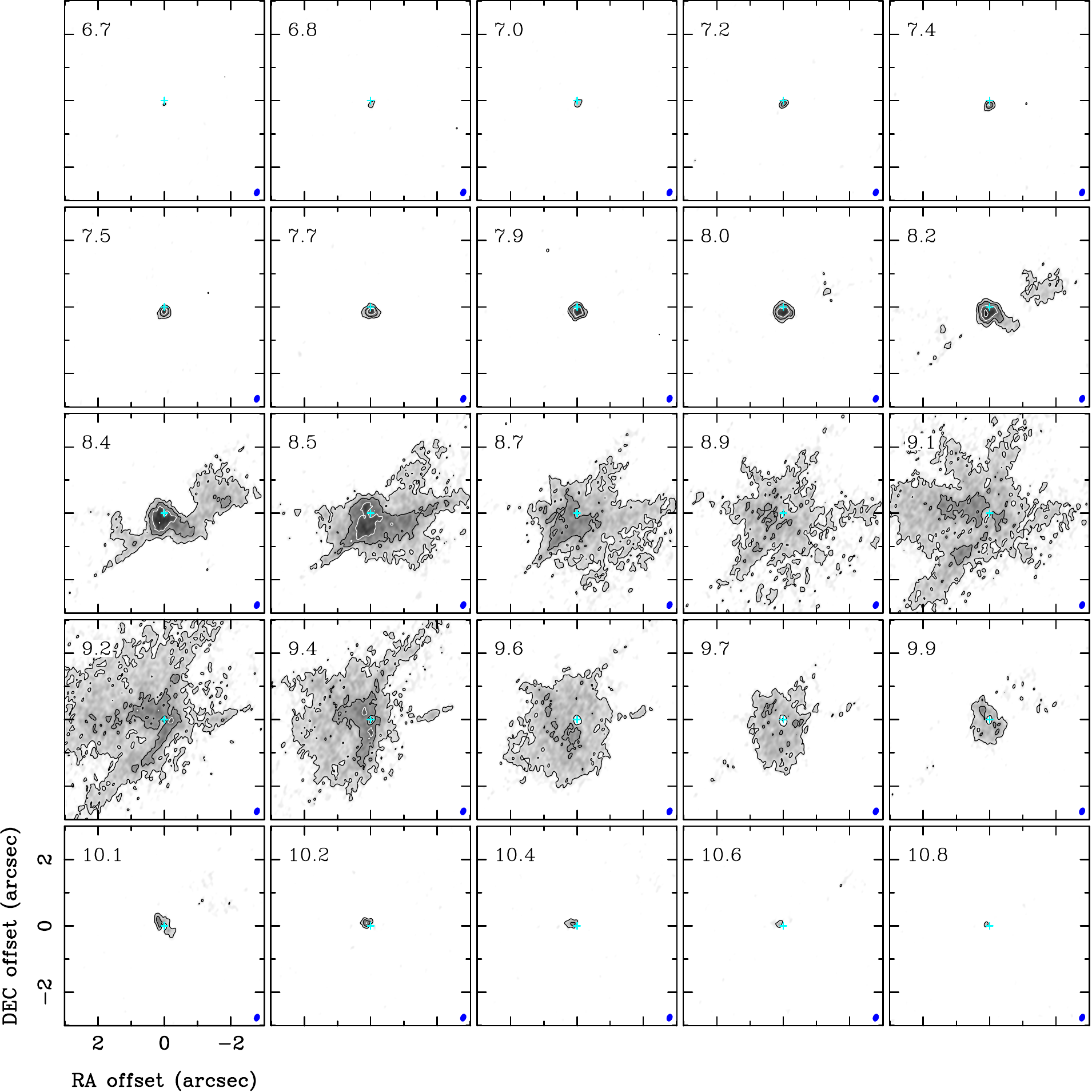}
\caption{Velocity channel maps of the C$^{18}$O (2--1) emission in HH~211 obtained with the ALMA observations. Filled ellipses present the size of the synthesized beam of 0\farcs2$\times$0\farcs13 (64 au $\times$ 42 au), and crosses show the protostellar position. Contours are from 5$\sigma$ in steps of 5$\sigma$, where 1$\sigma$ is 1.6 mJy beam$^{-1}$. Numbers in the upper left corners are the central LSR velocities of the channels in units of km s$^{-1}$.}\label{c18o_chans}
\end{figure*}

\begin{figure*}
\centering
\includegraphics[width=\textwidth]{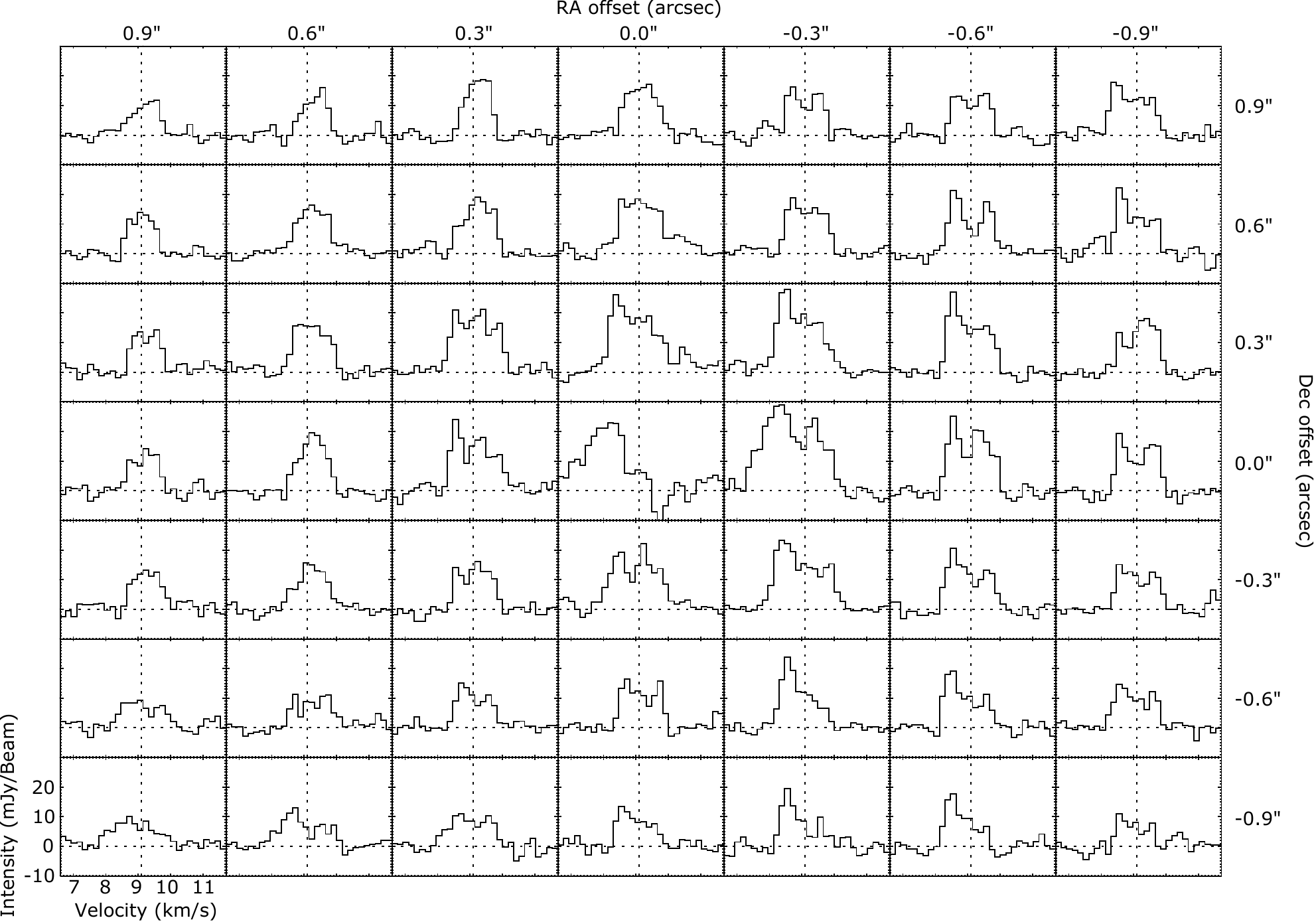}
\caption{Spectra of the C$^{18}$O (2--1) emission at different positions in HH~211 obtained with the ALMA observations. The spectra are extracted at offsets of ($-0\farcs9$, $-0\farcs9$) to ($0\farcs9$, $0\farcs9$) with respect to the protostellar position in steps of 0\farcs3 along right ascension and declination. Vertical dashed lines denote the systemic velocity of 9.1 km s$^{-1}$.}\label{c18o_spec}
\end{figure*}

Figure~\ref{c18o_chans} presents the velocity channel maps of the C$^{18}$O (2--1) emission in HH~211 obtained with the ALMA observations.
At relative velocities higher than 1 km s$^{-1}$ with respect to the systemic velocity of 9.1 km s$^{-1}$, 
the C$^{18}$O emission is compact with a size smaller than 0\farcs5 and exhibits a velocity gradient along the northeast--southwest direction, where the northeastern and southwestern parts are red- and blueshifted, respectively. 
The direction of this velocity gradient is the same as that observed in the N$_2$H$^+$ emission in the dense core.
At relative velocities lower than 0.5 km s$^{-1}$, the emission along the outflow cavity wall, which is in the northwest--southeast direction, is seen. 
In the velocity channel maps, we do not find the signs of asymmetric infalling flows, as seen in some other protostellar sources \citep{Yen14, Pineda20}. 
Figure~\ref{c18o_spec} shows the spectra of the C$^{18}$O (2--1) emission in HH~211 obtained with the ALMA observations. 
At outer radii of $\sim$0\farcs9, the spectra primarily show single velocity components with the intensity peaks are red- and blueshifted at northeastern and southwestern sides. 
In the region around the center, the spectra tend to show double peaks with a dip around the systemic velocity, 
and the blueshifted peak is brighter than or comparable to the redshifted peak, which is commonly seen in infalling protostellar envelopes \citep{Ohashi97}. 
The inverse P-Cygni is also seen in the spectrum at the center.
Our kinematical model suggests that the C$^{18}$O emission in this central 1$\arcsec$ region is mostly optically thick. 
A higher optical depth tends to cause a deeper dip around the systemic velocity in the spectrum of a protostellar envelope.
This is different from the cases of optically thin lines in protostellar envelopes, where the double peaked spectra and complex velocity fields may hint at non-isotropic infalling motions \citep{Cabedo21}.
We also note that extended emission is observed in the C$^{18}$O (3--2) line with JCMT in this region \citep{Curtis10}, so the low-velocity channels of our C$^{18}$O (2--1) data may have significant missing fluxes. 

\section{Corner plots of the model fitting to the ALMA continuum and C$^{18}$O data}\label{app_mcmc}
\begin{figure*}
\centering
\includegraphics[width=\textwidth]{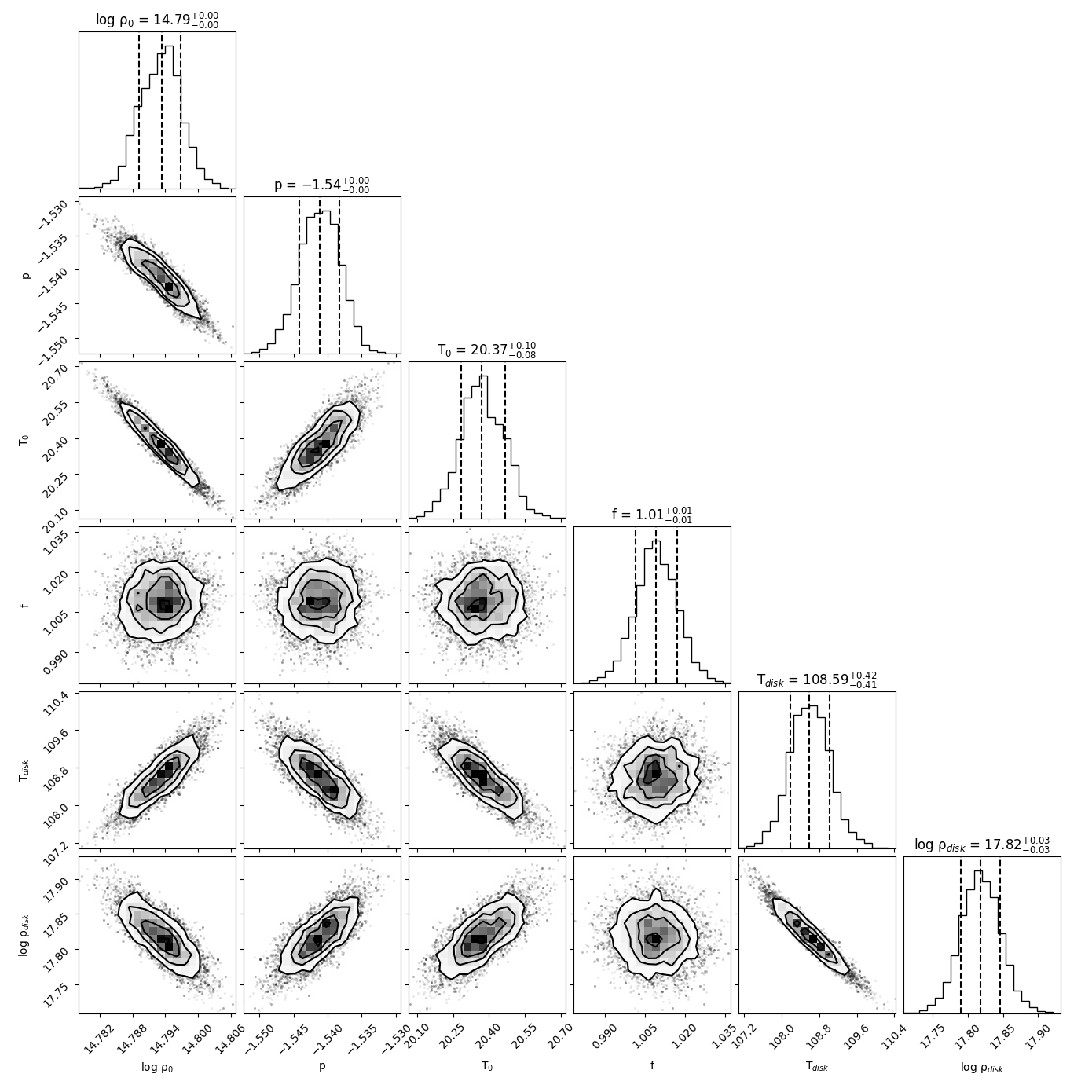}
\caption{Corner plots of our model fitting to the ALMA continuum data. Vertical dashed lines show the quantiles of 16\%, 50\% and 84\%.}\label{corner_cont}
\end{figure*}

\begin{figure*}
\centering
\includegraphics[width=\textwidth]{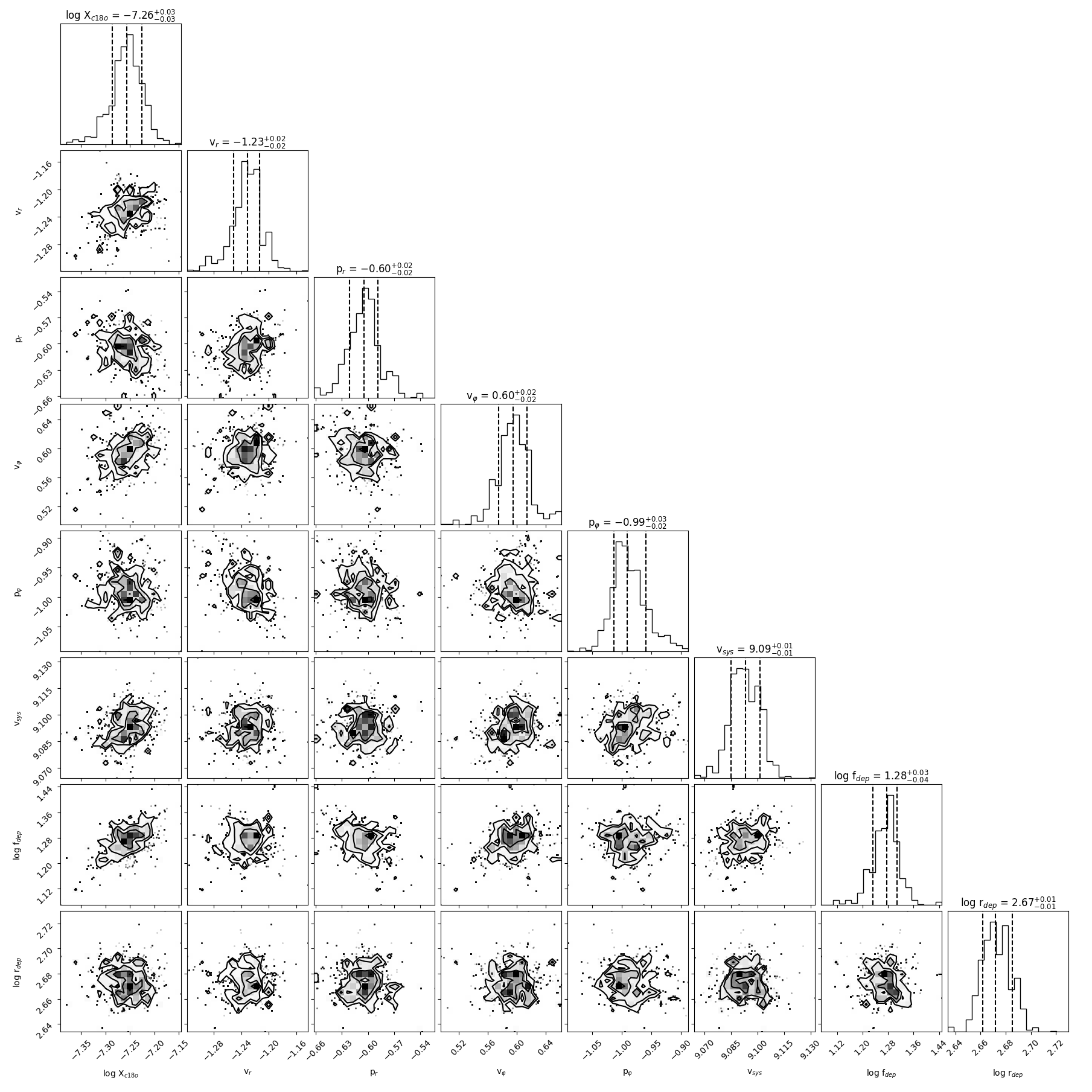}
\caption{Corner plots of our model fitting to the ALMA C$^{18}$O data. Vertical dashed lines show the quantiles of 16\%, 50\% and 84\%.}\label{corner_line}
\end{figure*}

Figure \ref{corner_cont} and \ref{corner_line} present the corner plots of our model fitting to the ALMA continuum and C$^{18}$O data, which show the correlations between the fitting parameters. 
The uncertainties from the model fitting are on the order of a few percentages, which are negligible compared to other uncertainties of more than 50\%, such as the magnetic field structures and dust absorption coefficients, in our estimate of the magnetic field strengths in the protostellar envelope (Section \ref{sec_env_b} and \ref{sec_dkappa}).

\end{appendix}

\software{NOSTAR \citep{Sawada08}, CASA \citep{McMullin07}, Starlink \citep{Currie14}, astrodendro \citep{Robitaille19}, emcee \citep{Foreman-Mackey13}, corner \citep{Foreman-Mackey16}}
\end{document}